\documentclass[12pt,english,12pt,sort&compress]{extarticle}
\usepackage{ae,aecompl}
\usepackage[T1]{fontenc}
\usepackage[latin9]{inputenc}
\usepackage{color}
\usepackage{amsmath}
\usepackage{graphicx}
\usepackage{esint}
\usepackage[numbers]{natbib}

\makeatletter
\usepackage{amsfonts}
\usepackage{aecompl}\usepackage{epstopdf}

\usepackage{babel}
\usepackage{babel}

\makeatother

\usepackage{babel}
\begin{document}

\title{Energy spectrum of a Langevin oscillator}

\author{Y.~Mishin$^{1,2}$~and J.~Hickman$^{1}$\bigskip{}
\\
{\normalsize{}$^{1}$ Department of Physics and Astronomy, George
Mason University, MSN 3F3,}\\
 {\normalsize{}Fairfax, VA 22030, USA}\\
{\normalsize{}}\\
$^{2}$ {\normalsize{}Corresponding author. E-mail address: ymishin@gmu.edu
(Y. Mishin)}}
\maketitle
\begin{abstract}
\textcolor{black}{We derive analytical solutions for the autocorrelation
and cross-correlation functions of the kinetic, potential and total
energy of a Langevin oscillator. These functions are presented in
both the time and frequency domains and validated by independent numerical
simulations. The results are applied to address the long-standing
issue of temperature fluctuations in canonical systems. }
\end{abstract}
\noindent \textcolor{black}{\emph{Keywords:}}\textcolor{black}{{} Langevin
equation, damped oscillator, energy spectrum, temperature fluctuation,
molecular dynamics}

\section{\textcolor{black}{Introduction\label{sec:Introduction}}}

\textcolor{black}{The Langevin equation is widely used for the modeling
of stochastic processes in many fields of physics and various branches
of science and engineering \citep{Coffey:2004aa}. In particular,
the equation can describe Brownian motion of a particle in a harmonic
potential well, often referred to as the Langevin oscillator. While
many properties of the Langevin oscillator have been exhaustively
studied over the past century, to our knowledge the correlation functions
and other statistical characteristics of the oscillator energy have
not been reported so far. }

\textcolor{black}{The goal of this paper is to investigate the fluctuations
of the kinetic, potential and total energy of a one-dimensional Langevin
oscillator. The results are presented in the form of analytical expressions
for the respective autocorrelation functions (ACFs) and cross-correlation
functions (CCFs) and their frequency spectra. The paper heavily relies
on the formalism of spectral representation of stochastic processes.
Some of the basic formalism is reviewed in Appendix A. The calculations
are enabled by a product rule of pair correlation functions presented
in Appendix B. The correlation functions reported in this work permit
a clear separation of two different timescales inherent in the Langevin
model. This timescale separation is a key to addressing the delicate,
and still controversial, issue of temperature fluctuations in systems
connected to a thermostat. }

\textcolor{black}{The Langevin equation for a one-dimensional harmonic
oscillator with a natural (resonant) frequency $\omega_{0}$ and a
friction coefficient (damping constant) $\gamma$ has the form \citep{Landau-Lifshitz-Stat-phys,Kubo:1966aa,Kubo:1991aa}
\begin{equation}
m\ddot{x}=-m\gamma v-m\omega_{0}^{2}x+R,\label{eq:1}
\end{equation}
where $m$ is the particle mass, $x$ is its deviation from equilibrium,
$v=\dot{x}$ is the velocity, and the random force (noise) $R$ satisfies
the condition $\bar{R}=0$. Here and everywhere below, the bar denotes
the canonical ensemble average. The variance of $R$ is adjusted to
balance the friction force and achieve equilibrium with the thermostat
at a chosen temperature $T_{0}$. The random force pumps mechanical
energy into the oscillator by incessant tiny kicks and causes thermal
fluctuations, whereas the friction force dissipates this energy into
heat. }

\textcolor{black}{Equation (\ref{eq:1}) is solved by spectral methods
\citep{Landau-Lifshitz-Stat-phys,Kubo:1966aa,Kubo:1991aa}. Taking
its Fourier transform we obtain
\begin{equation}
\hat{x}(\omega)=\dfrac{\hat{R}(\omega)/m}{\omega_{0}^{2}-\omega^{2}+i\gamma\omega},\label{eq:2}
\end{equation}
where the hat marks a Fourier transform with the angular frequency
$\omega$ (see Appendix A). For the particle velocity we have
\begin{equation}
\hat{v}(\omega)=i\omega\hat{x}(\omega)=\dfrac{i\omega\hat{R}(\omega)/m}{\omega_{0}^{2}-\omega^{2}+i\gamma\omega}.\label{eq:3}
\end{equation}
The random force $R$ is considered to be a white noise, for which
\begin{equation}
\overline{\hat{R}(\omega)\hat{R}(\omega^{\prime})}=\delta(\omega+\omega^{\prime})I_{R},\label{eq:6-1-1}
\end{equation}
where $I_{R}$ is a constant. Practically, this condition is satisfied
when the correlation time of $R$ is much shorter than both the vibration
period $2\pi/\omega_{0}$ and the damping time $1/\gamma$. The standard
calculation of $\overline{v^{2}}$ and application of the equipartition
theorem leads to the fluctuation-dissipation relation \citep{Landau-Lifshitz-Stat-phys,Kubo:1966aa,Kubo:1991aa}}

\textcolor{black}{
\begin{equation}
I_{R}=\dfrac{\gamma mkT_{0}}{\pi}\label{eq:17}
\end{equation}
linking the noise power $I_{R}$ to the damping constant $\gamma$.}

\textcolor{black}{The Fourier transform $\hat{C}_{xx}(\omega)$ of
the position ACF $C_{xx}(t)=\overline{x(0)x(t)}$ is obtained by inserting
$\hat{x}(\omega)$ from Eq.(\ref{eq:2}) into Eq.(\ref{eq:A28}) (Wiener-Khinchin
theorem, Appendix A):
\begin{eqnarray}
\overline{\hat{x}(\omega)\hat{x}(\omega^{\prime})} & = & \dfrac{\overline{\hat{R}(\omega)\hat{R}(\omega^{\prime})}/m^{2}}{(\omega_{0}^{2}-\omega^{2}+i\gamma\omega)(\omega_{0}^{2}-\omega^{\prime2}+i\gamma\omega^{\prime})}\nonumber \\
 & = & \dfrac{(I_{R}/m^{2})\delta(\omega+\omega^{\prime})}{(\omega_{0}^{2}-\omega^{2}+i\gamma\omega)(\omega_{0}^{2}-\omega^{\prime2}+i\gamma\omega^{\prime})}=\hat{C}_{xx}(\omega)\delta(\omega+\omega^{\prime}),\label{eq:6-2}
\end{eqnarray}
where
\begin{equation}
\hat{C}_{xx}(\omega)=\dfrac{\gamma kT_{0}/\pi m}{\left(\omega_{0}^{2}-\omega^{2}\right)^{2}+\gamma^{2}\omega^{2}}.\label{eq:17-2}
\end{equation}
A similar calculation gives the spectral form of the velocity ACF:
\begin{equation}
\hat{C}_{vv}(\omega)=\dfrac{(\gamma kT_{0}/\pi m)\omega^{2}}{\left(\omega_{0}^{2}-\omega^{2}\right)^{2}+\gamma^{2}\omega^{2}},\label{eq:17-1}
\end{equation}
where we used Eq.(\ref{eq:3}) for $\hat{v}(\omega)$. The position-velocity
CCF $\hat{C}_{xv}(\omega)$ is obtained in a similar manner using
Eq.(\ref{eq:A19}) from Appendix A:}

\textcolor{black}{
\begin{equation}
\hat{C}_{xv}(\omega)=-\dfrac{i(\gamma kT_{0}/\pi m)\omega}{\left(\omega_{0}^{2}-\omega^{2}\right)^{2}+\gamma^{2}\omega^{2}}.\label{eq:24-2}
\end{equation}
}

\textcolor{black}{The correlation functions (\ref{eq:17-2}), (\ref{eq:17-1})
and (\ref{eq:24-2}) are well-known and are only reproduced here as
ingredients for the subsequent calculations. }

\section{\textcolor{black}{Kinetic energy of the Langevin oscillator\label{sec:KE}}}

\textcolor{black}{Our goal is to compute the ACF $C_{\Delta K\Delta K}(t)=\overline{\Delta K(0)\Delta K(t)}$
of the kinetic energy $K=mv^{2}/2$ relative to its average value
$\overline{K}=kT_{0}/2$, where we denote $\Delta K=K-\overline{K}$.
We first find the spectral ACF $\hat{C}_{\Delta K\Delta K}(\omega)$
by applying the equations derived in Appendix B. Taking $a(t)=v(t)$,
Eq.(\ref{eq:B9}) gives 
\begin{equation}
\hat{C}_{\Delta K\Delta K}(\omega)=\dfrac{m^{2}}{2}\intop_{-\infty}^{\infty}\hat{C}_{vv}(\omega^{\prime})\hat{C}_{vv}(\omega-\omega^{\prime})d\omega^{\prime}.\label{eq:32}
\end{equation}
Inserting $\hat{C}_{vv}(\omega)$ from Eq.(\ref{eq:17-1}) we have
\begin{equation}
\hat{C}_{\Delta K\Delta K}(\omega)=\dfrac{(\gamma kT_{0})^{2}}{2\pi^{2}}\intop_{-\infty}^{\infty}\dfrac{\omega^{\prime2}(\omega-\omega^{\prime})^{2}d\omega^{\prime}}{\left[\left(\omega_{0}^{2}-\omega^{\prime2}\right)^{2}+\gamma^{2}\omega^{\prime2}\right]\left[\left(\omega_{0}^{2}-(\omega-\omega^{\prime})^{2}\right)^{2}+\gamma^{2}(\omega-\omega^{\prime})^{2}\right]}.\label{eq:33}
\end{equation}
}

\begin{figure}
\noindent \begin{centering}
\textbf{\textcolor{black}{(a)}}\textcolor{black}{\qquad{}\includegraphics[scale=0.75]{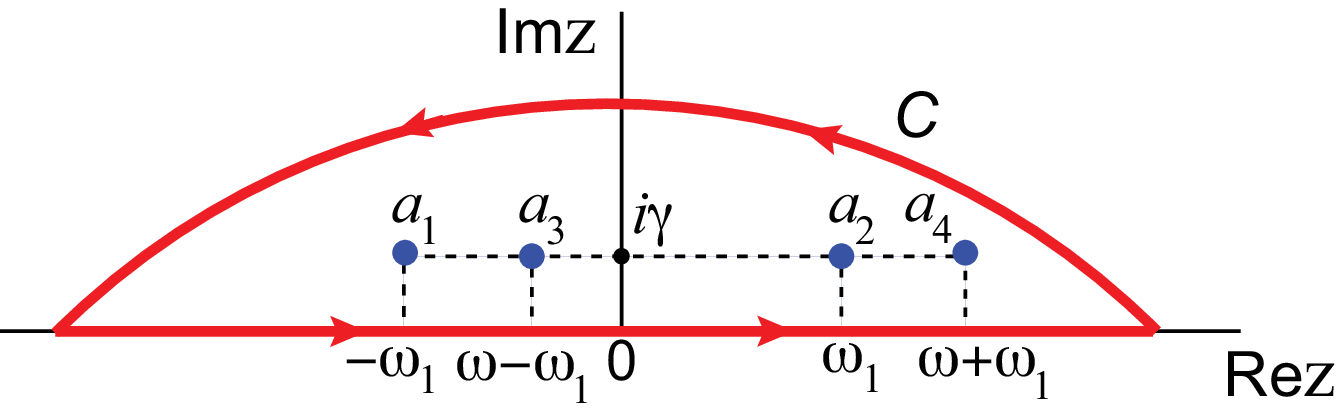}}
\par\end{centering}
\textcolor{black}{\bigskip{}
}
\noindent \begin{centering}
\textbf{\textcolor{black}{(b)}}\textcolor{black}{\qquad{}\includegraphics[scale=0.75]{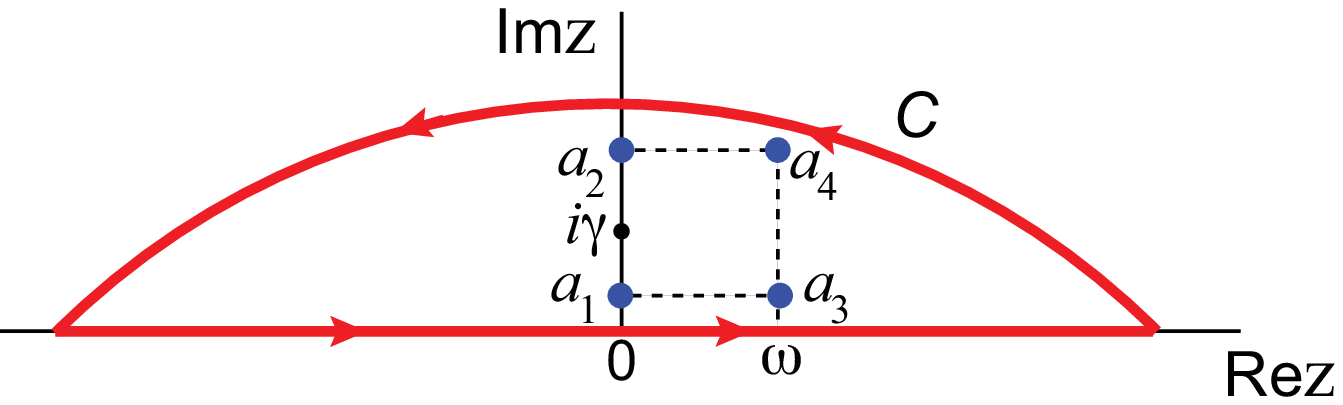}}
\par\end{centering}
\textcolor{black}{\caption{Integration loop for computing $\hat{C}_{\Delta K\Delta K}(\omega)$
using Eq.(\ref{eq:33}). The four singularity points enclosed by the
loop are indicated. (a) $\omega_{0}>\gamma/2$, (b) $\omega_{0}<\gamma/2$.
\label{fig:loop2}}
}
\end{figure}

\begin{figure}
\noindent \begin{centering}
\textbf{\textcolor{black}{(a)}}\textcolor{black}{\qquad{}\includegraphics[scale=0.77]{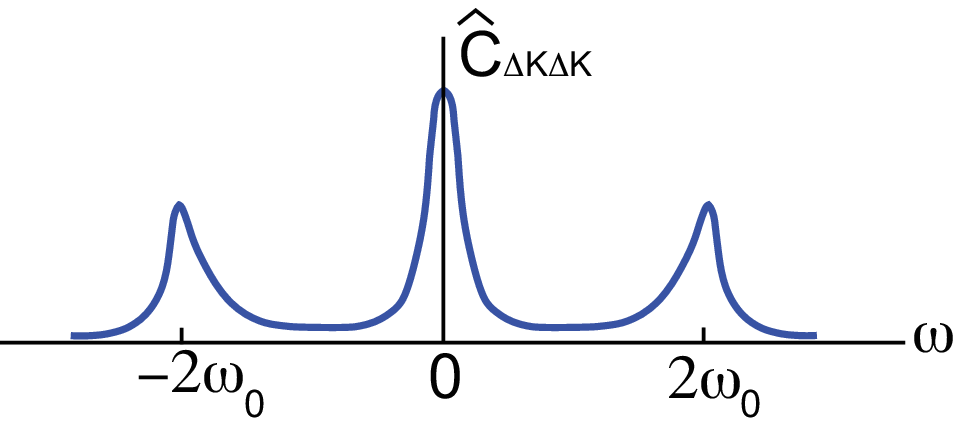}}
\par\end{centering}
\noindent \begin{centering}
\textbf{\textcolor{black}{(b)}}\textcolor{black}{\qquad{}\includegraphics[scale=0.77]{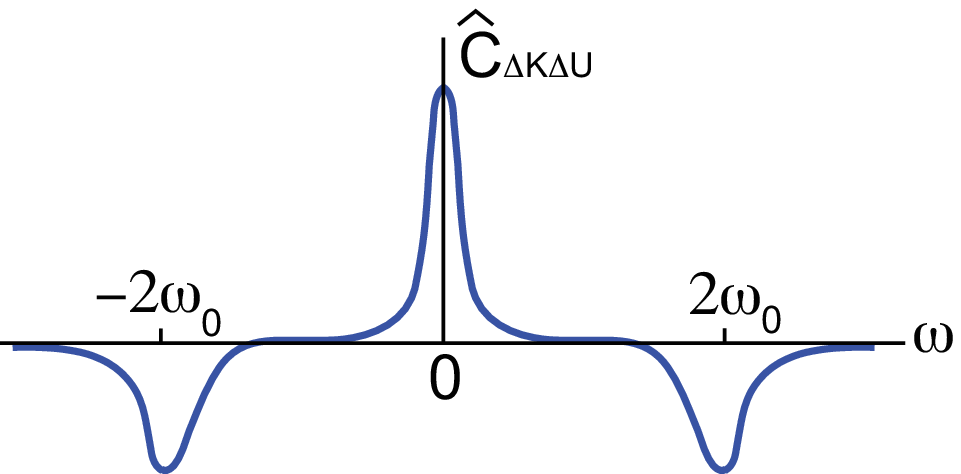}}
\par\end{centering}
\textcolor{black}{\caption{Schematic plots of the kinetic energy auto-correlation function (a)
and kinetic-potential energy cross-correlation function (b) of a Langevin
oscillator in the frequency domain.\label{fig:2} }
}
\end{figure}

\textcolor{black}{The integral in Eq.(\ref{eq:33}) is evaluated by
replacing $\omega^{\prime}$ by a complex variable $z$ and integrating
the function
\begin{equation}
f(z)=\dfrac{z^{2}(\omega-z)^{2}}{\left[\left(\omega_{0}^{2}-z^{2}\right)^{2}+\gamma^{2}z^{2}\right]\left[\left(\omega_{0}^{2}-(\omega-z)^{2}\right)^{2}+\gamma^{2}(\omega-z)^{2}\right]}\label{eq:34}
\end{equation}
along a semi-circular closed loop $C$ in the complex plane (Fig.\ \ref{fig:loop2}(a)).
This function has eight singularities, the following four of which
lie inside the loop:
\begin{equation}
a_{1}=-\omega_{1}+i\gamma/2,\:\:a_{2}=\omega_{1}+i\gamma/2,\:\:a_{3}=\omega-\omega_{1}+i\gamma/2,\:\:a_{4}=\omega+\omega_{1}+i\gamma/2,\label{eq:11-2}
\end{equation}
where
\begin{equation}
\omega_{1}=\sqrt{\omega_{0}^{2}-\gamma^{2}/4}\label{eq:12}
\end{equation}
and we assumed that $\omega_{0}>\gamma/2$. Finding the residues at
these singularities and inserting them in the residue theorem we have
\begin{equation}
\ointop_{C}f(z)dz=2\pi i\sum_{i=1}^{4}\mathrm{Res}(f,a_{i})=2\pi\dfrac{\omega^{4}+\omega^{2}(\gamma^{2}-3\omega_{0}^{2})+4\omega_{0}^{4}}{\gamma(\gamma^{2}+\omega^{2})\left[4\gamma^{2}\omega^{2}+(\omega^{2}-4\omega_{0}^{2})^{2}\right]}.\label{eq:35}
\end{equation}
If $\omega_{0}<\gamma/2$, the four singularities enclosed by the
loop are (Fig.\ \ref{fig:loop2}(b)): 
\begin{equation}
a_{1}=-i\omega_{2}+i\gamma/2,\:\:a_{2}=i\omega_{2}+i\gamma/2,\:\:a_{3}=\omega-i\omega_{2}+i\gamma/2,\:\:a_{4}=\omega+i\omega_{2}+i\gamma/2,\label{eq:11-2-1}
\end{equation}
where
\begin{equation}
\omega_{2}=\sqrt{\gamma^{2}/4-\omega_{0}^{2}},\label{eq:12-1}
\end{equation}
and the calculations give the same result as in Eq.(\ref{eq:35}).
When $|z|\rightarrow\infty$, $|f(z)|$ tends to zero as $1/|z|^{4}$
and the integral along the arc vanishes, leaving only the integral
from $-\infty$ to $\infty$ along the real axis appearing in Eq.(\ref{eq:33}).
We finally obtain the spectral ACF of the kinetic energy: 
\begin{equation}
\hat{C}_{\Delta K\Delta K}(\omega)=\dfrac{\gamma(kT_{0})^{2}}{\pi}\dfrac{\omega^{4}+\omega^{2}(\gamma^{2}-3\omega_{0}^{2})+4\omega_{0}^{4}}{(\gamma^{2}+\omega^{2})\left[4\gamma^{2}\omega^{2}+(\omega^{2}-4\omega_{0}^{2})^{2}\right]}.\label{eq:36}
\end{equation}
This function has three maxima: one at $\omega=0$ and two more near
$\omega=\pm2\omega_{0}$ (Fig.\ \ref{fig:2}(a)). }

\textcolor{black}{If $\gamma\ll\omega_{0}$ (underdamped regime),
the maxima of $\hat{C}_{\Delta K\Delta K}(\omega)$ are very sharp
and separated by frequency gaps. Near the central maximum we have
$|\omega|\ll\omega_{0}$ and Eq.(\ref{eq:36}) gives a Lorentz peak
of width $\gamma$: 
\begin{equation}
\hat{C}_{\Delta K\Delta K}(\omega)=\dfrac{\gamma(kT_{0})^{2}}{4\pi}\dfrac{1}{\gamma^{2}+\omega^{2}}\label{eq:38}
\end{equation}
The ACF corresponding to this peak is
\begin{equation}
C_{\Delta K\Delta K}(t)=\intop_{-\infty}^{\infty}\hat{C}_{\Delta K\Delta K}(\omega)e^{i\omega t}d\omega=\dfrac{\gamma(kT_{0})^{2}}{4\pi}\intop_{-\infty}^{\infty}\dfrac{e^{i\omega t}}{\gamma^{2}+\omega^{2}}d\omega.\label{eq:39}
\end{equation}
The integral is readily computed using the residue theorem with the
same semi-circular loop as before. The loop encloses one singularity
at $a=i\gamma$. The integral along the arc vanishes and we obtain
(assuming $t>0$)
\begin{equation}
C_{\Delta K\Delta K}(t)=\dfrac{(kT_{0})^{2}}{4}e^{-\gamma t}.\label{eq:40}
\end{equation}
This function describes long-range fluctuations of $K$ due to energy
exchanges with the thermostat. }

\textcolor{black}{Around the remaining maxima we have $|\omega\pm2\omega_{0}|\ll\omega_{0}$
and $\hat{C}_{\Delta K\Delta K}(\omega)$ can be approximated by 
\begin{equation}
\hat{C}_{\Delta K\Delta K}(\omega)=\dfrac{\gamma(kT_{0})^{2}}{8\pi}\dfrac{1}{\gamma^{2}+(\omega\pm2\omega_{0})^{2}}.\label{eq:41}
\end{equation}
This is again a Lorentz function of width $\gamma$, except that the
height of these peaks is half of that at $\omega=0$. The ACF corresponding
to these peaks is found by inverse Fourier transformation:
\begin{equation}
C_{\Delta K\Delta K}(t)=\dfrac{\gamma(kT_{0})^{2}}{8\pi}\intop_{-\infty}^{\infty}\dfrac{e^{i\omega t}}{\gamma^{2}+(\omega\pm2\omega_{0})^{2}}d\omega.\label{eq:43}
\end{equation}
We again apply the residue theorem using the same integration loop.
There are two singularities lying inside the loop: $a_{\pm}=\pm2\omega_{0}+i\gamma$
and we obtain (assuming $t>0$) 
\begin{equation}
C_{\Delta K\Delta K}(t)=\dfrac{(kT_{0})^{2}}{4}e^{-\gamma t}\cos(2\omega_{0}t).\label{eq:42}
\end{equation}
The peak at $\omega=2\omega_{0}$ describes the kinetic energy variations
during quasi-harmonic oscillations. Since the kinetic and potential
energies transform to each other twice per each period, the frequency
is $2\omega_{0}$. The peak at $\omega=-2\omega_{0}$ describes physically
the same process and only appears in the spectrum to formally satisfy
the definition of the Fourier transformation.}

\textcolor{black}{The general form of $C_{\Delta K\Delta K}(t)$ is
obtained by inverse Fourier transformation of Eq.(\ref{eq:36}):
\begin{equation}
C_{\Delta K\Delta K}(t)=\dfrac{\gamma(kT_{0})^{2}}{\pi}\intop_{-\infty}^{\infty}\dfrac{\left[\omega^{4}+\omega^{2}(\gamma^{2}-3\omega_{0}^{2})+4\omega_{0}^{4}\right]e^{i\omega t}}{(\gamma^{2}+\omega^{2})\left[4\gamma^{2}\omega^{2}+(\omega^{2}-4\omega_{0}^{2})^{2}\right]}d\omega.\label{eq:43-1}
\end{equation}
As usual, we apply the residue theorem. Suppose $\omega_{0}>\gamma/2$.
Then the function
\[
f(z)=\dfrac{\left[z^{4}+z^{2}(\gamma^{2}-3\omega_{0}^{2})+4\omega_{0}^{4}\right]e^{izt}}{(\gamma^{2}+z^{2})\left[4\gamma^{2}z^{2}+(z^{2}-4\omega_{0}^{2})^{2}\right]}
\]
has three singularities in the upper half-plane ($\textnormal{Im}z>0$):}

\begin{figure}
\noindent \begin{centering}
\textcolor{black}{\includegraphics[scale=0.75]{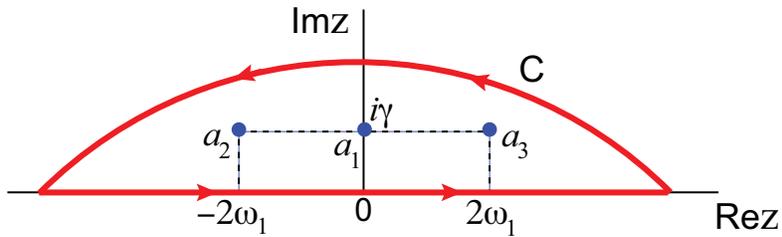}}
\par\end{centering}
\textcolor{black}{\caption{Integration loop for computing $\hat{C}_{KK}(t)$ using Eq.(\ref{eq:43-1}).
The singularity points lying inside the loop when $\omega_{0}>\gamma/2$
are indicated.\label{fig:loop3}}
}
\end{figure}

\textcolor{black}{
\begin{equation}
a_{1}=i\gamma,\:\:a_{2}=i\gamma-2\omega_{1},\:\:a_{3}=i\gamma+2\omega_{1},\label{eq:11-2-1-1}
\end{equation}
with $\omega_{1}$ given by Eq.(\ref{eq:12}). Choosing the same semi-circular
integration path as before (Fig.~\ref{fig:loop3}), we have 
\begin{equation}
\ointop_{C}f(z)dz=2\pi i\sum_{i=1}^{3}\mathrm{Res}(f,a_{i})=\dfrac{\pi e^{-\gamma t}}{8\gamma\omega_{1}^{2}}\left[2\omega_{0}^{2}+(2\omega_{0}^{2}-\gamma^{2})\cos(2\omega_{1}t)-2\gamma\omega_{1}\sin(2\omega_{1}t)\right].\label{eq:43-2}
\end{equation}
The integral along the arc vanishes and we finally obtain
\begin{equation}
C_{\Delta K\Delta K}(t)=\dfrac{(kT_{0})^{2}e^{-\gamma t}}{8\omega_{1}^{2}}\left[2\omega_{0}^{2}+(2\omega_{0}^{2}-\gamma^{2})\cos(2\omega_{1}t)-2\gamma\omega_{1}\sin(2\omega_{1}t)\right].\label{eq:43-3}
\end{equation}
If $\omega_{0}<\gamma/2$, then similar calculations give
\begin{equation}
C_{\Delta K\Delta K}(t)=-\dfrac{(kT_{0})^{2}e^{-\gamma t}}{8\omega_{2}^{2}}\left[2\omega_{0}^{2}+(2\omega_{0}^{2}-\gamma^{2})\cosh(2\omega_{2}t)-2\gamma\omega_{2}\sinh(2\omega_{2}t)\right],\label{eq:43-4}
\end{equation}
where $\omega_{2}$ is given by Eq.(\ref{eq:12-1}). In the latter
case, all three singularities lie on the imaginary axis.}

\textcolor{black}{Knowing $C_{\Delta K\Delta K}(t)$ we can find the
mean-square fluctuation $\overline{(\Delta K)^{2}}=C_{\Delta K\Delta K}(0)$.
Eqs.(\ref{eq:43-3}) and (\ref{eq:43-4}) both give the same result:
$\overline{(\Delta K)^{2}}=(kT_{0})^{2}/2$, which matches the independent
calculation from the canonical distribution. On the other hand, using
Eqs.(\ref{eq:40}) and (\ref{eq:42}), we find that the peaks at $\omega=0$
and $\omega=2\omega_{0}$ make equal contributions $\overline{(\Delta K)^{2}}=(kT_{0})^{2}/4$.
Thus, one half of the kinetic energy fluctuation $\overline{(\Delta K)^{2}}$
is caused by quasi-harmonic vibrations, whereas the other half is
due to energy fluctuations between the oscillator and the thermostat. }

\section{\textcolor{black}{Potential energy of the Langevin oscillator\label{sec:PE}}}

\textcolor{black}{We next calculate the ACF $C_{\Delta U\Delta U}(t)$
of the potential energy $U=m\omega_{0}^{2}x^{2}/2$ relative to its
average value $\overline{U}=kT_{0}/2$, where $\Delta U=U-\overline{U}$.
As with kinetic energy, we first find the spectral ACF $\hat{C}_{\Delta U\Delta U}(\omega)$
using the approximation discussed in Appendix B with $a(t)=x(t)$.
Applying Eq.(\ref{eq:B9}) we have 
\begin{equation}
\hat{C}_{\Delta U\Delta U}(\omega)=\dfrac{m^{2}\omega_{0}^{4}}{2}\intop_{-\infty}^{\infty}\hat{C}_{xx}(\omega^{\prime})\hat{C}_{xx}(\omega-\omega^{\prime})d\omega^{\prime}.\label{eq:45}
\end{equation}
Inserting $\hat{C}_{xx}(\omega)$ from Eq.(\ref{eq:17-2}),
\begin{equation}
\hat{C}_{\Delta U\Delta U}(\omega)=\dfrac{(\gamma kT_{0})^{2}\omega_{0}^{4}}{2\pi^{2}}\intop_{-\infty}^{\infty}\dfrac{d\omega^{\prime}}{\left[\left(\omega_{0}^{2}-\omega^{\prime2}\right)^{2}+\gamma^{2}\omega^{\prime2}\right]\left[\left(\omega_{0}^{2}-(\omega-\omega^{\prime})^{2}\right)^{2}+\gamma^{2}(\omega-\omega^{\prime})^{2}\right]}.\label{eq:46}
\end{equation}
The right-hand side is evaluated by integrating the complex function
\begin{equation}
f(z)=\dfrac{1}{\left[\left(\omega_{0}^{2}-z^{2}\right)^{2}+\gamma^{2}z^{2}\right]\left[\left(\omega_{0}^{2}-(\omega-z)^{2}\right)^{2}+\gamma^{2}(\omega-z)^{2}\right]}\label{eq:54}
\end{equation}
along a semi-circular loop $C$ in the complex plane (Fig.\ \ref{fig:loop2}).
The loop encloses the same four singularities as for the kinetic energy.
The residue theorem gives
\begin{equation}
\ointop_{C}f(z)dz=2\pi i\sum_{i=1}^{4}\mathrm{Res}(f,a_{i})=2\pi\dfrac{4\gamma^{2}+\omega^{2}+4\omega_{0}^{2}}{\gamma\omega_{0}^{2}(\gamma^{2}+\omega^{2})\left[4\gamma^{2}\omega^{2}+(\omega^{2}-4\omega_{0}^{2})^{2}\right]},\label{eq:56}
\end{equation}
from which 
\begin{equation}
\hat{C}_{\Delta U\Delta U}(\omega)=\dfrac{\gamma(kT_{0})^{2}\omega_{0}^{2}}{\pi}\dfrac{4\gamma^{2}+\omega^{2}+4\omega_{0}^{2}}{(\gamma^{2}+\omega^{2})\left[4\gamma^{2}\omega^{2}+(\omega^{2}-4\omega_{0}^{2})^{2}\right]}.\label{eq:57}
\end{equation}
Similar to the kinetic energy case, this function has maxima at $\omega=0$
and near $\omega=\pm2\omega_{0}$, which have the same physical meaning:
the maximum at $\omega=0$ describes long-range fluctuations due to
energy exchanges with the thermostat, whereas the maximum near $\omega=2\omega_{0}$
is due to quasi-harmonic vibrations. Again, the maximum near $-2\omega_{0}$
represents physically the same process; the formal negative frequencies
are only shown on the plots to better visualize the central peak.}

\textcolor{black}{The real-time ACF $C_{\Delta U\Delta U}(t)$ is
calculated by inverse Fourier transformation of Eq.(\ref{eq:57}):
\begin{equation}
C_{\Delta U\Delta U}(t)=\intop_{-\infty}^{\infty}\hat{C}_{\Delta U\Delta U}(\omega)e^{i\omega t}d\omega=\dfrac{\gamma(kT_{0})^{2}\omega_{0}^{2}}{\pi}\intop_{-\infty}^{\infty}\dfrac{\left(4\gamma^{2}+\omega^{2}+4\omega_{0}^{2}\right)e^{i\omega t}}{(\gamma^{2}+\omega^{2})\left[4\gamma^{2}\omega^{2}+(\omega^{2}-4\omega_{0}^{2})^{2}\right]}d\omega.\label{eq:58}
\end{equation}
The function
\begin{equation}
f(z)=\dfrac{\left[4\gamma^{2}+z^{2}+4\omega_{0}^{2}\right]e^{izt}}{(\gamma^{2}+z^{2})\left[4\gamma^{2}z^{2}+(z^{2}-4\omega_{0}^{2})^{2}\right]}\label{eq:59}
\end{equation}
has the same three singularities in the upper half-plane ($\textnormal{Im}z>0$)
as its kinetic energy counterpart (Fig.~\ref{fig:loop3}). Choosing
the same semi-circular integration path and assuming that $\omega_{0}>\gamma/2$,
the residue theorem gives 
\begin{equation}
\ointop_{C}f(z)dz=2\pi i\sum_{i=1}^{3}\mathrm{Res}(f,a_{i})=\dfrac{\pi e^{-\gamma t}}{8\omega_{0}^{2}\omega_{1}^{2}}\left[2\omega_{0}^{2}+(2\omega_{0}^{2}-\gamma^{2})\cos(2\omega_{1}t)+2\gamma\omega_{1}\sin(2\omega_{1}t)\right].\label{eq:60}
\end{equation}
The integral along the arc vanishes and we obtain
\begin{equation}
C_{\Delta U\Delta U}(t)=\dfrac{(kT_{0})^{2}e^{-\gamma t}}{8\omega_{1}^{2}}\left[2\omega_{0}^{2}+(2\omega_{0}^{2}-\gamma^{2})\cos(2\omega_{1}t)+2\gamma\omega_{1}\sin(2\omega_{1}t)\right].\label{eq:61}
\end{equation}
When $\omega_{0}<\gamma/2$, similar calculations give
\begin{equation}
C_{\Delta U\Delta U}(t)=-\dfrac{(kT_{0})^{2}e^{-\gamma t}}{8\omega_{2}^{2}}\left[2\omega_{0}^{2}+(2\omega_{0}^{2}-\gamma^{2})\cosh(2\omega_{2}t)+2\gamma\omega_{2}\sinh(2\omega_{2}t)\right].\label{eq:62}
\end{equation}
Note that $C_{\Delta U\Delta U}(t)$ looks similar but is different
from the previously derived $C_{\Delta K\Delta K}(t)$.}

\textcolor{black}{Knowing $C_{\Delta U\Delta U}(t)$, we find $\overline{(\Delta U)^{2}}=C_{\Delta U\Delta U}(0)=(kT_{0})^{2}/2$.
In the strongly underdamped (quasi-harmonic) regime, this fluctuation
is split equally between quasi-harmonic vibrations and energy exchanges
with the thermostat. }

\section{\textcolor{black}{Total energy of the Langevin oscillator\label{sec:E}}}

\textcolor{black}{The total energy of the oscillator can be factorized
as follows:
\begin{equation}
E=\dfrac{mv^{2}}{2}+\dfrac{m\omega_{0}^{2}x^{2}}{2}=\dfrac{m}{2}ab,\label{eq:63}
\end{equation}
where
\begin{equation}
a\equiv v+i\omega_{0}x,\:\:b\equiv v-i\omega_{0}x.\label{eq:64}
\end{equation}
To find the ACF $C_{\Delta E\Delta E}(t)$ (where $\Delta E=E-\overline{E}$),
we first calculate the spectral form of this ACF. Using the equation
from Appendix B, }

\textcolor{black}{
\begin{equation}
\hat{C}_{\Delta E\Delta E}(\omega)=\dfrac{m^{2}}{4}\intop_{-\infty}^{\infty}\hat{C}_{aa}(\omega^{\prime})\hat{C}_{bb}(\omega-\omega^{\prime})d\omega^{\prime}+\dfrac{m^{2}}{4}\intop_{-\infty}^{\infty}\hat{C}_{ba}(\omega^{\prime})\hat{C}_{ab}(\omega-\omega^{\prime})d\omega^{\prime}.\label{eq:65}
\end{equation}
}

\textcolor{black}{The correlation functions appearing in Eq.(\ref{eq:65})
are computed as follows. We have
\begin{eqnarray}
\overline{\hat{a}(\omega)\hat{a}(\omega^{\prime})} & = & \overline{\hat{v}(\omega)\hat{v}(\omega^{\prime})}-\omega_{0}^{2}\overline{\hat{x}(\omega)\hat{x}(\omega^{\prime})}+i\omega_{0}\overline{\hat{x}(\omega)\hat{v}(\omega^{\prime})}+i\omega_{0}\overline{\hat{v}(\omega)\hat{x}(\omega^{\prime})}\nonumber \\
 & = & \delta(\omega+\omega^{\prime})\left[\hat{C}_{vv}(\omega)-\omega_{0}^{2}\hat{C}_{xx}(\omega)+i\omega_{0}\hat{C}_{xv}(-\omega)+i\omega_{0}\hat{C}_{vx}(-\omega)\right],\label{eq:66}
\end{eqnarray}
where we used Eqs.(\ref{eq:A19}) and (\ref{eq:A28}) from Appendix
A. The last two terms cancel each other and we obtain
\begin{equation}
\overline{\hat{a}(\omega)\hat{a}(\omega^{\prime})}=\delta(\omega+\omega^{\prime})\left[\hat{C}_{vv}(\omega)-\omega_{0}^{2}\hat{C}_{xx}(\omega)\right],\label{eq:67}
\end{equation}
from which
\begin{equation}
\hat{C}_{aa}(\omega)=\hat{C}_{vv}(\omega)-\omega_{0}^{2}\hat{C}_{xx}(\omega).\label{eq:68}
\end{equation}
Similar calculations give
\begin{equation}
\hat{C}_{bb}(\omega)=\hat{C}_{vv}(\omega)-\omega_{0}^{2}\hat{C}_{xx}(\omega).\label{eq:69}
\end{equation}
For the cross-correlation $\hat{C}_{ab}$ we have
\begin{eqnarray}
\overline{\hat{a}(\omega)\hat{b}(\omega^{\prime})} & = & \overline{\hat{v}(\omega)\hat{v}(\omega^{\prime})}+\omega_{0}^{2}\overline{\hat{x}(\omega)\hat{x}(\omega^{\prime})}+i\omega_{0}\overline{\hat{x}(\omega)\hat{v}(\omega^{\prime})}-i\omega_{0}\overline{\hat{v}(\omega)\hat{x}(\omega^{\prime})}\nonumber \\
 & = & \delta(\omega+\omega^{\prime})\left[\hat{C}_{vv}(\omega)+\omega_{0}^{2}\hat{C}_{xx}(\omega)+i\omega_{0}\hat{C}_{xv}(-\omega)-i\omega_{0}\hat{C}_{vx}(-\omega)\right],\label{eq:70}
\end{eqnarray}
from which
\begin{equation}
\hat{C}_{ab}(\omega)=\omega_{0}^{2}\hat{C}_{xx}(\omega)+\hat{C}_{vv}(\omega)-2i\omega_{0}\hat{C}_{vx}(\omega).\label{eq:71}
\end{equation}
The functions $\hat{C}_{vv}(\omega)$, $\hat{C}_{xx}(\omega)$ and
$\hat{C}_{vx}(\omega)$ are given by Eqs.(\ref{eq:17-1}), (\ref{eq:17-2})
and (\ref{eq:24-2}), respectively. Inserting them in Eqs.(\ref{eq:68}),
(\ref{eq:69}) and (\ref{eq:71}) we obtain
\begin{equation}
\hat{C}_{aa}(\omega)=\hat{C}_{bb}(\omega)=\dfrac{(\gamma kT_{0}/\pi m)(\omega^{2}-\omega_{0}^{2})}{\left(\omega_{0}^{2}-\omega^{2}\right)^{2}+\gamma^{2}\omega^{2}},\label{eq:72}
\end{equation}
\begin{equation}
\hat{C}_{ab}(\omega)=\hat{C}_{ba}(-\omega)=\dfrac{(\gamma kT_{0}/\pi m)(\omega-\omega_{0})^{2}}{\left(\omega_{0}^{2}-\omega^{2}\right)^{2}+\gamma^{2}\omega^{2}}.\label{eq:73}
\end{equation}
These functions provide the input to Eq.(\ref{eq:65}), which then
becomes
\begin{eqnarray}
\hat{C}_{\Delta E\Delta E}(\omega) & = & \dfrac{(\gamma kT_{0})^{2}}{4\pi^{2}}\intop_{-\infty}^{\infty}\dfrac{(\omega_{0}+\omega^{\prime})^{2}(\omega-\omega^{\prime}-\omega_{0})^{2}d\omega^{\prime}}{\left[\left(\omega_{0}^{2}-\omega^{\prime2}\right)^{2}+\gamma^{2}\omega^{\prime2}\right]\left[\left(\omega_{0}^{2}-(\omega-\omega^{\prime})^{2}\right)^{2}+\gamma^{2}(\omega-\omega^{\prime})^{2}\right]}d\omega^{\prime}\nonumber \\
 & + & \dfrac{(\gamma kT_{0})^{2}}{4\pi^{2}}\intop_{-\infty}^{\infty}\dfrac{(\omega_{0}+\omega^{\prime})^{2}(\omega-\omega^{\prime}-\omega_{0})^{2}d\omega^{\prime}}{\left[\left(\omega_{0}^{2}-\omega^{\prime2}\right)^{2}+\gamma^{2}\omega^{\prime2}\right]\left[\left(\omega_{0}^{2}-(\omega-\omega^{\prime})^{2}\right)^{2}+\gamma^{2}(\omega-\omega^{\prime})^{2}\right]}d\omega^{\prime}.\label{eq:74}
\end{eqnarray}
}

\textcolor{black}{The integrals are readily evaluated using the residue
theorem with the same semi-circular integration loop as for the kinetic
and potential energies. The singularities of the integrands lying
inside the loop are the same as in Eqs.(\ref{eq:33}) and (\ref{eq:46}).
Somewhat lengthy calculations give}

\textcolor{black}{
\begin{equation}
\hat{C}_{\Delta E\Delta E}(\omega)=\dfrac{\gamma(kT_{0})^{2}}{\pi}\dfrac{(\omega^{2}-4\omega_{0}^{2})^{2}+\gamma^{2}(\omega^{2}+4\omega_{0}^{2})}{(\gamma^{2}+\omega^{2})\left[4\gamma^{2}\omega^{2}+(\omega^{2}-4\omega_{0}^{2})^{2}\right]}.\label{eq:75}
\end{equation}
This function has a maximum at $\omega=0$ and local minima near $\pm2\omega_{0}$.
When $\gamma\ll\omega_{0}$, these extrema are separated by frequency
gaps. Near the maximum, $\hat{C}_{\Delta E\Delta E}(\omega)$ behaves
as 
\begin{equation}
\hat{C}_{\Delta E\Delta E}(\omega)=\dfrac{\gamma(kT_{0})^{2}}{\pi}\dfrac{1}{\gamma^{2}+\omega^{2}}.\label{eq:76}
\end{equation}
This is a Lorentz peak of width $\gamma$ and height $(kT_{0})^{2}/\pi\gamma$.
This peak represents the energy fluctuations between the system and
the thermostat and is four times as high as the similar peaks for
the kinetic and potential energies. Near $\omega=2\omega_{0}$, $\hat{C}_{\Delta E\Delta E}(\omega)$
behaves approximately as 
\begin{equation}
\hat{C}_{\Delta E\Delta E}(\omega)=\dfrac{\gamma(kT_{0})^{2}}{8\pi\omega_{0}^{2}}\dfrac{\gamma^{2}+2(\omega-2\omega_{0})^{2}}{\gamma^{2}+(\omega-2\omega_{0})^{2}}.\label{eq:77}
\end{equation}
This equation describes a Lorentz-shape local minimum of width $\gamma$
and depth $\gamma(kT_{0})^{2}/8\pi\omega_{0}^{2}$. This depth is
a factor of $\gamma^{2}/8\omega_{0}^{2}$ smaller than the height
of the maximum $\omega=0$. In the strongly underdamped regime ($\gamma\ll\omega_{0}$),
this minimum is extremely shallow and can be neglected. It describes
an ``anti-resonance'' effect wherein the oscillator is less willing
to exchange the total energy with the thermostat at the natural frequency
of the kinetic-potential energy fluctuations (which is $2\omega_{0}$)
than at nearly frequencies. In the underdamped regime this is a tiny
second-order effect. Most of the energy exchanges between the oscillator
and the thermostat occur at low frequencies. }

\textcolor{black}{The time-dependent ACF $C_{\Delta E\Delta E}(t)$
can now be obtained by inverse Fourier transformation of Eq.(\ref{eq:75}):
\begin{equation}
C_{\Delta E\Delta E}(t)=\intop_{-\infty}^{\infty}\hat{C}_{\Delta E\Delta E}(\omega)e^{i\omega t}d\omega=\dfrac{\gamma(kT_{0})^{2}}{\pi}\intop_{-\infty}^{\infty}\dfrac{\left[(\omega^{2}-4\omega_{0}^{2})^{2}+\gamma^{2}(\omega^{2}+4\omega_{0}^{2})\right]e^{i\omega t}}{(\gamma^{2}+\omega^{2})\left[4\gamma^{2}\omega^{2}+(\omega^{2}-4\omega_{0}^{2})^{2}\right]}d\omega.\label{eq:78}
\end{equation}
As before, we apply the residue theorem utilizing the semi-circular
integration loop shown in Fig.~\ref{fig:loop3}. We obtain
\begin{equation}
C_{\Delta E\Delta E}(t)=\dfrac{(kT_{0})^{2}}{4\omega_{1}^{2}}e^{-\gamma t}\left[4\omega_{0}^{2}-\gamma^{2}\cos(2\omega_{1}t)\right]\label{eq:79}
\end{equation}
if $\omega_{0}>\gamma/2$ and 
\begin{equation}
C_{\Delta E\Delta E}(t)=-\dfrac{(kT_{0})^{2}}{4\omega_{2}^{2}}e^{-\gamma t}\left[4\omega_{0}^{2}-\gamma^{2}\cosh(2\omega_{2}t)\right]\label{eq:79-1}
\end{equation}
if $\omega_{0}<\gamma/2$. These equations correctly give the mean-square
fluctuation of the total energy:
\begin{equation}
\overline{(\Delta E)^{2}}=(kT_{0})^{2}.\label{eq:80}
\end{equation}
}

\section{\textcolor{black}{The cross-correlation functions}}

\textcolor{black}{In this section we calculate the CCFs between the
kinetic, potential and total energies. We start by computing the spectral
form of the kinetic-potential energy CCF $\hat{C}_{\Delta K\Delta U}(\omega)$
using the equations from Appendix B with $a(t)=v(t)$ and $b(t)=x(t)$.
In the notations of Appendix B, $G(t)=v^{2}(t)$ and $H(t)=x^{2}(t)$.
Equation (\ref{eq:B7-1-1}) gives
\begin{eqnarray}
\hat{C}_{\Delta K\Delta U}(\omega) & = & \dfrac{m^{2}\omega_{0}^{2}}{4}\intop_{-\infty}^{\infty}\hat{C}_{xv}(\omega^{\prime})\hat{C}_{xv}(\omega-\omega^{\prime})d\omega^{\prime}\nonumber \\
 & = & -\intop_{-\infty}^{\infty}\dfrac{(\gamma kT\omega_{0}/4\pi)^{2}\omega^{\prime}(\omega-\omega^{\prime})d\omega^{\prime}}{\left[\left(\omega_{0}^{2}-\omega^{\prime2}\right)^{2}+\gamma^{2}\omega^{\prime2}\right]\left[\left(\omega_{0}^{2}-(\omega-\omega^{\prime})^{2}\right)^{2}+\gamma^{2}(\omega-\omega^{\prime})^{2}\right]}d\omega^{\prime}.\label{eq:81}
\end{eqnarray}
At the second step we inserted $\hat{C}_{xv}(\omega)$ from Eq.(\ref{eq:24-2}).
The integral is evaluated by integration along the usual path $C$
in the complex plane (Fig.\ \ref{fig:loop2}). The loop contains
the same singularities as in the ACF calculations for the kinetic
and potential energies. Calculations employing the residue theorem
give 
\begin{equation}
\hat{C}_{\Delta K\Delta U}(\omega)=\dfrac{\gamma(kT_{0})^{2}}{\pi}\dfrac{\omega_{0}^{2}(4\omega_{0}^{2}-3\omega^{2})}{(\gamma^{2}+\omega^{2})\left[4\gamma^{2}\omega^{2}+(\omega^{2}-4\omega_{0}^{2})^{2}\right]}.\label{eq:82}
\end{equation}
This function has a central maximum at $\omega=0$ and two negative
minima at $\omega=\pm2\omega_{0}$ (Fig.~\ref{fig:2}(b)). When $\gamma\ll\omega_{0}$,
these extrema are separated by frequency gaps and have a Lorentz shape
of width $\gamma$ and the heights of $(kT_{0})^{2}/4\pi\gamma$ and
$-(kT_{0})^{2}/8\pi\gamma$, respectively. As before, the central
maximum represents the energy exchanges with the thermostat while
the minima arise from quasi-harmonic vibrations. The negative sign
of the minima reflects the fact that the kinetic and potential energies
oscillate in anti-phase: when one increases, the other decreases. }

\textcolor{black}{Since $\hat{C}_{\Delta K\Delta U}(\omega)$ is an
even function of frequency, $\hat{C}_{\Delta U\Delta K}(\omega)$
is given by the same equation (\ref{eq:82}). We can now calculate
the CCFs of the total energy with the kinetic and potential energies.
We have
\begin{equation}
\hat{C}_{\Delta E\Delta K}(\omega)=\hat{C}_{\Delta K\Delta K}(\omega)+\hat{C}_{\Delta U\Delta K}(\omega)=\dfrac{\gamma(kT_{0})^{2}}{\pi}\dfrac{\omega^{4}+(\gamma^{2}-6\omega_{0}^{2})\omega^{2}+8\omega_{0}^{4}}{(\gamma^{2}+\omega^{2})\left[4\gamma^{2}\omega^{2}+(\omega^{2}-4\omega_{0}^{2})^{2}\right]},\label{eq:83}
\end{equation}
where we used Eq.(\ref{eq:36}) for $\hat{C}_{\Delta K\Delta K}(\omega)$.
Similarly,
\begin{equation}
\hat{C}_{\Delta E\Delta U}(\omega)=\hat{C}_{\Delta K\Delta U}(\omega)+\hat{C}_{\Delta U\Delta U}(\omega)=\dfrac{\gamma(kT_{0})^{2}}{\pi}\dfrac{8\omega_{0}^{4}+4\gamma^{2}\omega_{0}^{2}-2\omega^{2}\omega_{0}^{2}}{(\gamma^{2}+\omega^{2})\left[4\gamma^{2}\omega^{2}+(\omega^{2}-4\omega_{0}^{2})^{2}\right]},\label{eq:84}
\end{equation}
where we used Eq.(\ref{eq:57}) for $\hat{C}_{\Delta U\Delta U}(\omega)$.
At $\gamma\ll\omega_{0}$, both $\hat{C}_{\Delta E\Delta K}(\omega)$
and $\hat{C}_{\Delta E\Delta U}(\omega)$ have a central peak at $\omega=0$
and a tiny wiggle near $\omega=\pm2\omega_{0}$, the latter being
associated with the ``anti-resonance'' effect mentioned above. Thus,
at low frequencies the kinetic and potential energies strongly correlate
with the total energy, which is consistent with the picture of long-range
fluctuations due to slow energy exchanges with the thermostat maintaining
nearly equilibrium partitioning between the kinetic and potential
energies. }

\textcolor{black}{The time domain forms of these CCFs are obtained
by Fourier transformations using the residue theorem and the semi-circular
integration path shown in Fig.~\ref{fig:loop3}. In all cases, the
three singularities enclosed by the path are given by Eq.(\ref{eq:11-2-1-1}).
The calculations are similar to those for the ACFs and, assuming $\omega_{0}>\gamma/2$,
give
\begin{equation}
C_{\Delta K\Delta U}(t)=\dfrac{(kT_{0})^{2}\omega_{0}^{2}}{4\omega_{1}^{2}}e^{-\gamma t}\left[1-\cos(2\omega_{1}t)\right],\label{eq:85}
\end{equation}
\begin{equation}
C_{\Delta E\Delta K}(t)=\dfrac{(kT_{0})^{2}e^{-\gamma t}}{8\omega_{1}^{2}}\left[4\omega_{0}^{2}-\gamma^{2}\cos(2\omega_{1}t)-2\gamma\omega_{1}\sin(2\omega_{1}t)\right],\label{eq:86}
\end{equation}
\begin{equation}
C_{\Delta E\Delta U}(t)=\dfrac{(kT_{0})^{2}e^{-\gamma t}}{8\omega_{1}^{2}}\left[4\omega_{0}^{2}-\gamma^{2}\cos(2\omega_{1}t)+2\gamma\omega_{1}\sin(2\omega_{1}t)\right].\label{eq:87}
\end{equation}
If $\omega_{0}<\gamma/2$, these equations become, respectively, 
\begin{equation}
C_{\Delta K\Delta U}(t)=-\dfrac{(kT_{0})^{2}\omega_{0}^{2}}{4\omega_{2}^{2}}e^{-\gamma t}\left[1-\cosh(2\omega_{2}t)\right],\label{eq:88}
\end{equation}
\begin{equation}
C_{\Delta E\Delta K}(t)=-\dfrac{(kT_{0})^{2}e^{-\gamma t}}{8\omega_{2}^{2}}\left[4\omega_{0}^{2}-\gamma^{2}\cosh(2\omega_{2}t)-2\gamma\omega_{1}\sinh(2\omega_{2}t)\right],\label{eq:89}
\end{equation}
\begin{equation}
C_{\Delta E\Delta U}(t)=-\dfrac{(kT_{0})^{2}e^{-\gamma t}}{8\omega_{2}^{2}}\left[4\omega_{0}^{2}-\gamma^{2}\cosh(2\omega_{2}t)+2\gamma\omega_{1}\sinh(2\omega_{2}t)\right].\label{eq:90}
\end{equation}
At $t=0$, these equations give $\overline{\Delta K\Delta U}=0$ and
$\overline{\Delta E\Delta K}=\overline{\Delta E\Delta U}=(kT_{0})^{2}/2$.}

\section{\textcolor{black}{Molecular dynamics simulations}}

\textcolor{black}{The analytical calculations presented in the previous
sections rely on the approximation discussed in Appendix B. In this
approximation, the four-member correlation functions are replaced
by sums of products of pair correlation functions. To demonstrate
the accuracy of this approximation, the energy ACFs and CCFs were
computed by molecular dynamics (MD) simulations and the results were
compared with the analytical solutions.}

\textcolor{black}{The Langevin equation (\ref{eq:1}) was integrated
numerically by implementing the velocity Verlet algorithm with $m=1$,
$\omega_{0}=1$ and $\gamma=0.1\omega_{0}$. Because $\gamma/\omega_{0}=0.1$
is relatively small, the simulations realize the underdamped regime.
The time step of integration was 0.001. Every 100 MD steps, the random
force $R$ was updated by drawing a new number from the normal distribution
with the standard deviation of 0.5. Alternatively, a uniform distributions
of $R$ was used in a few test runs and the same results were obtained.
(In fact, the popular LAMMPS molecular dynamics package \citep{Plimpton95}
implements the Langevin thermostat with a uniform distribution for
speed.) A total of 5000 statistically independent MD runs, each $80\gamma^{-1}$
long, were performed to reach convergence. For each run, the discrete
Fourier transformations of the kinetic, potential and total energies
were computed and the Fourier amplitudes were averaged over all MD
runs. The Fourier amplitudes obtained were used to calculate the respective
correlation functions in the frequency domain, which were then mapped
into the time domain by inverse Fourier transformation.}

\textcolor{black}{To facilitate comparison with the analytical solutions,
all correlation functions were expressed in terms of the dimensionless
frequency $\omega/\omega_{0}$, time $t\gamma$ and damping constant
$\gamma/\omega_{0}$, and normalized as follows
\begin{equation}
\hat{\mathcal{C}}_{\Delta X\Delta Y}(\omega/\omega_{0},\gamma/\omega_{0})=\dfrac{\hat{C}_{\Delta X\Delta Y}(\omega)}{\left(\overline{(\Delta X)^{2}}\enskip\overline{(\Delta Y)^{2}}\right)^{1/2}},\label{eq:97}
\end{equation}
\begin{equation}
\mathcal{C}_{\Delta X\Delta Y}(t\gamma,\gamma/\omega_{0})=\dfrac{C_{\Delta X\Delta Y}(t)}{\left(\overline{(\Delta X)^{2}}\enskip\overline{(\Delta Y)^{2}}\right)^{1/2}},\label{eq:98}
\end{equation}
where $X$ and $Y$ stand for $K$, $U$ or $E$, with $X\neq Y$
for CCFs and $X=Y$ for ACFs.}

\begin{figure}
\noindent \begin{centering}
\textbf{\textcolor{black}{(a)}}\textcolor{black}{\quad{}\includegraphics[width=0.5\textwidth]{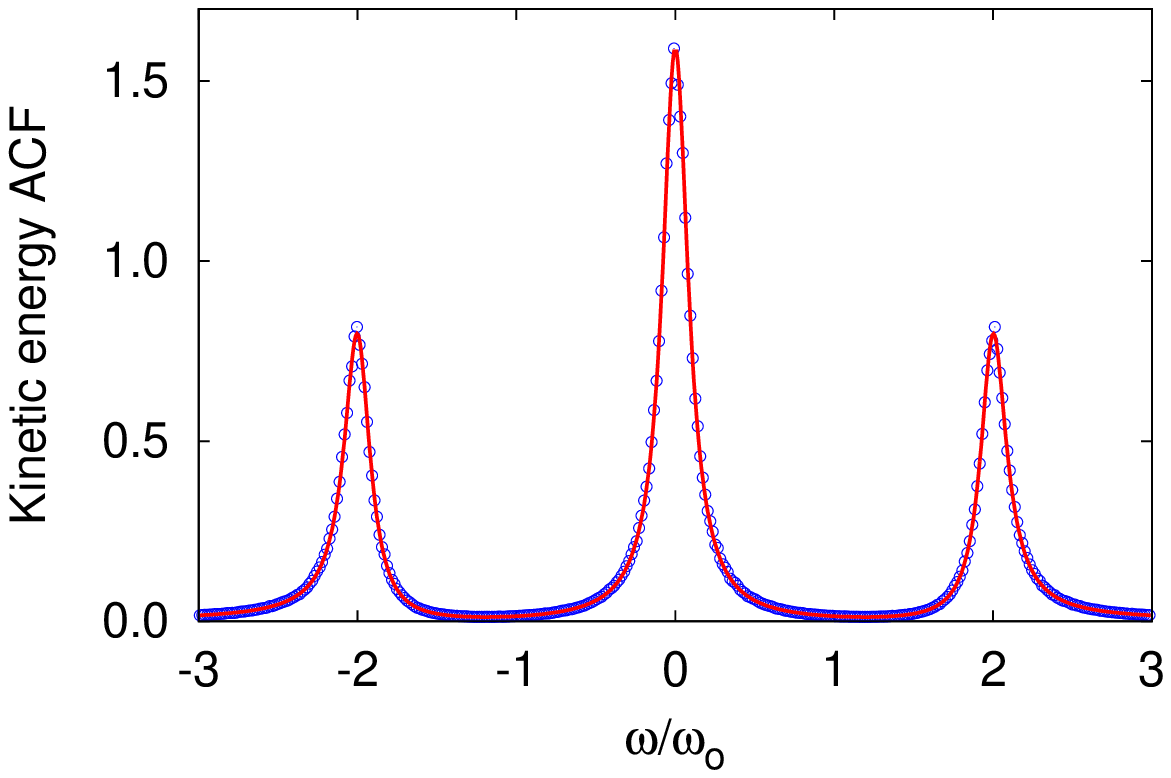}}
\par\end{centering}
\noindent \begin{centering}
\textbf{\textcolor{black}{(b)}}\textcolor{black}{\quad{}\includegraphics[width=0.5\textwidth]{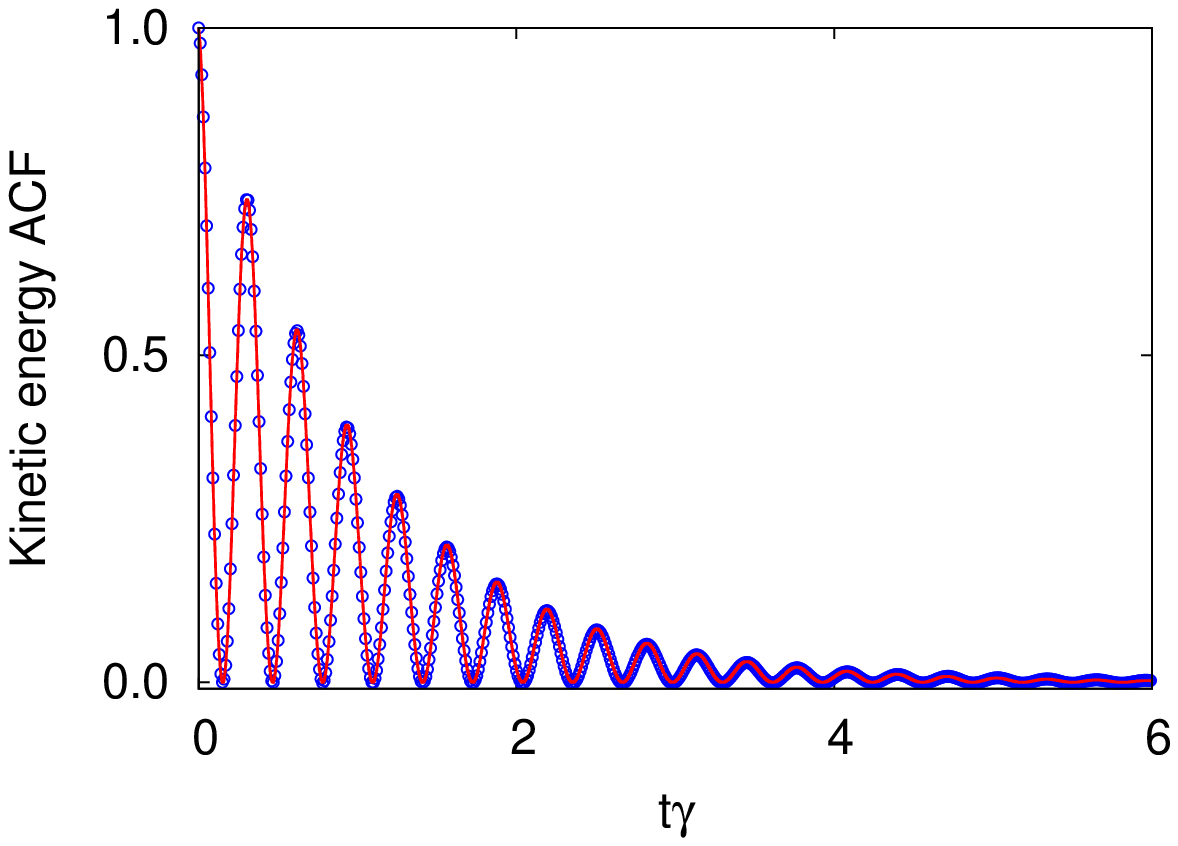}}
\par\end{centering}
\textcolor{black}{\caption{The kinetic energy ACF in the frequency (a) and time (b) domains for
underdamped vibrations with the damping constant $\gamma=0.1\omega_{0}$.
The points and lines represent MD results and analytical solutions,
respectively. \textcolor{black}{The functions are normalized according
to Eqs.(\ref{eq:97}) and (\ref{eq:98})}\textcolor{green}{.}\label{fig:K-ACF}}
}
\end{figure}

\begin{figure}
\noindent \begin{centering}
\textbf{\textcolor{black}{(a)}}\textcolor{black}{\quad{}\includegraphics[width=0.5\textwidth]{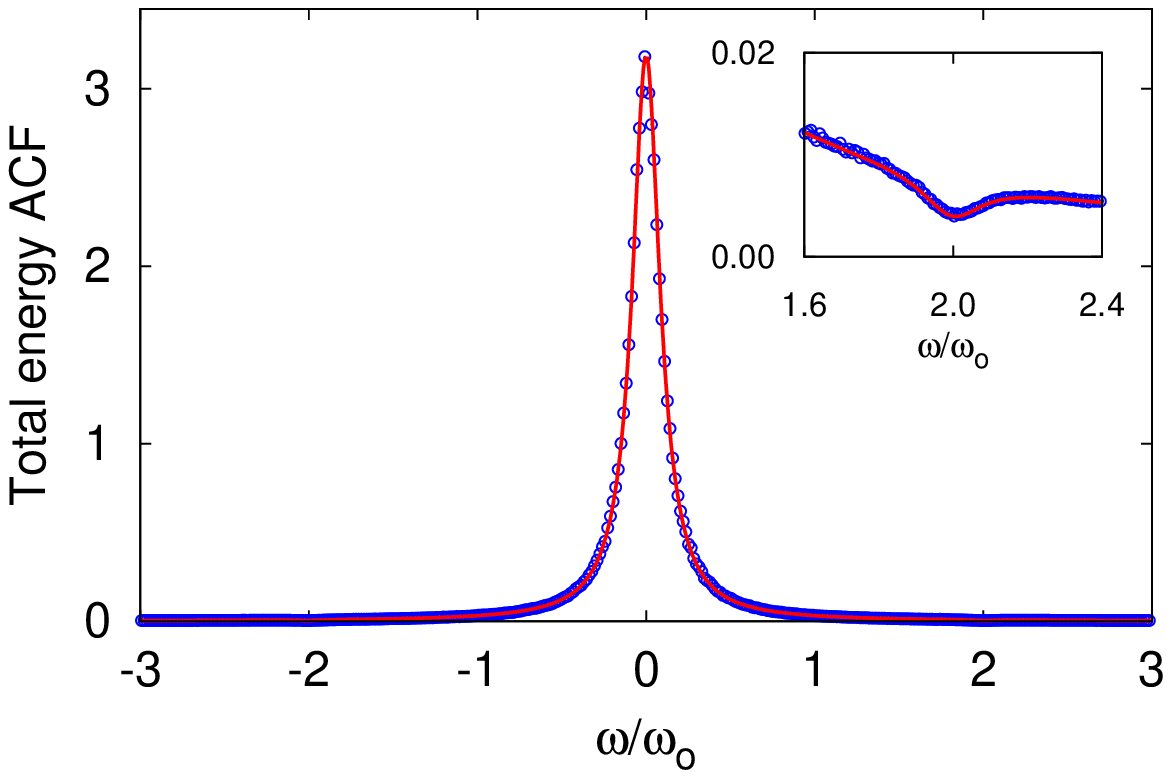}}
\par\end{centering}
\noindent \begin{centering}
\textbf{\textcolor{black}{(b)}}\textcolor{black}{\quad{}\includegraphics[width=0.5\textwidth]{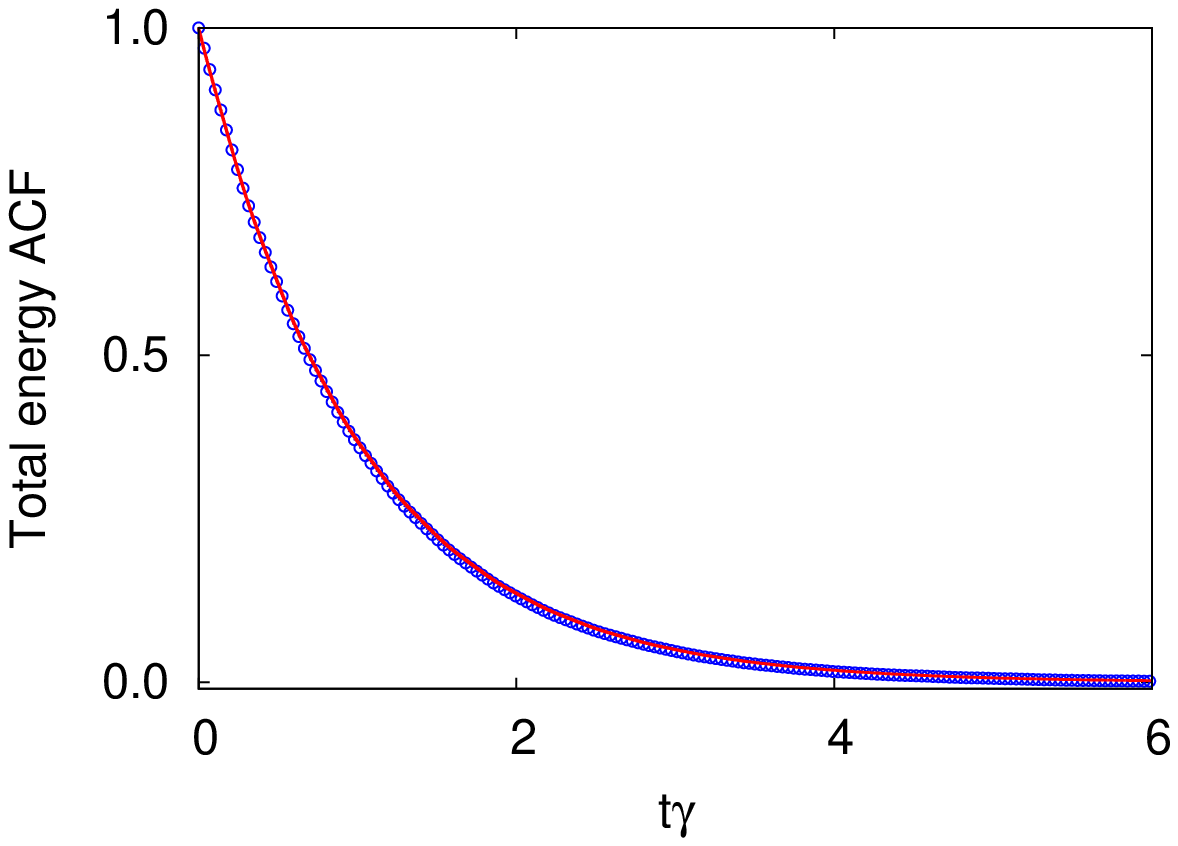}}
\par\end{centering}
\textcolor{black}{\caption{The total energy ACF in the frequency (a) and time (b) domains for
underdamped vibrations with the damping constant $\gamma=0.1\omega_{0}$.
The points and lines represent MD results and analytical solutions,
respectively. The inset shows a zoom into the ``anti-resonance'' region.
\textcolor{black}{The functions are normalized according to Eqs.(\ref{eq:97})
and (\ref{eq:98}).} \label{fig:E-ACF}}
}
\end{figure}

\begin{figure}
\noindent \begin{centering}
\textbf{\textcolor{black}{(a)}}\textcolor{black}{\quad{}\includegraphics[width=0.5\textwidth]{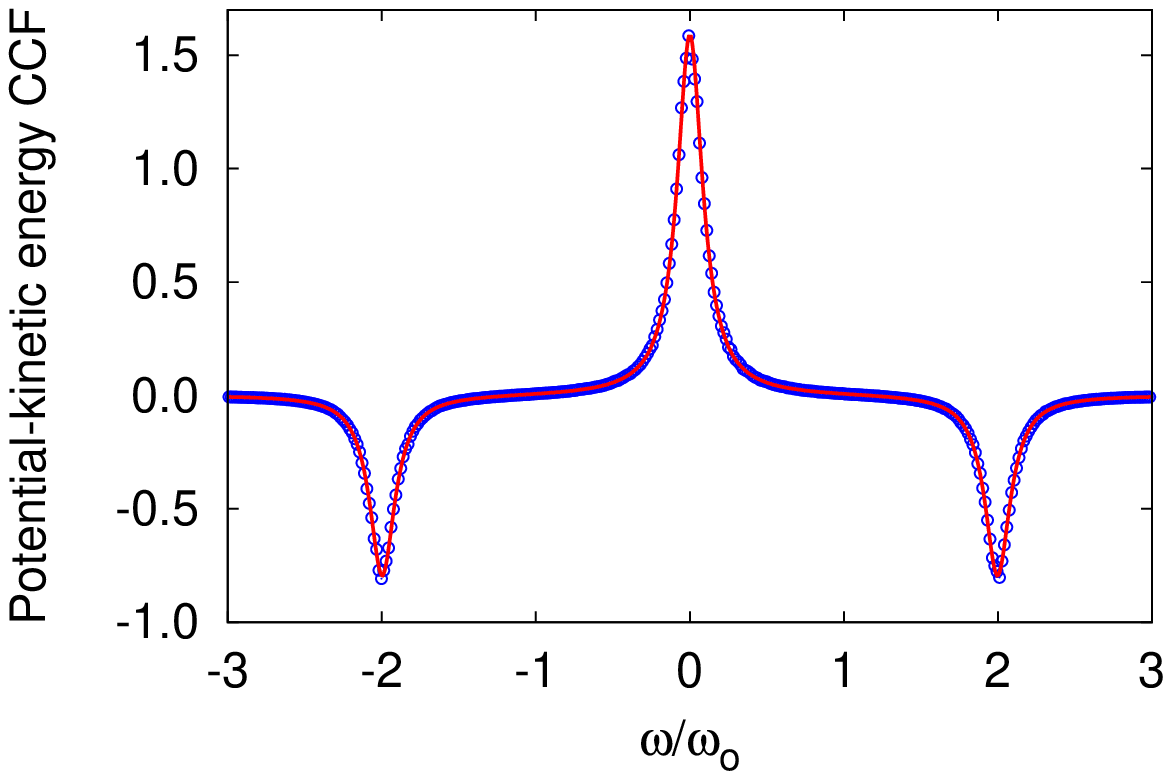}}
\par\end{centering}
\noindent \begin{centering}
\textbf{\textcolor{black}{(b)}}\textcolor{black}{\quad{}\includegraphics[width=0.5\textwidth]{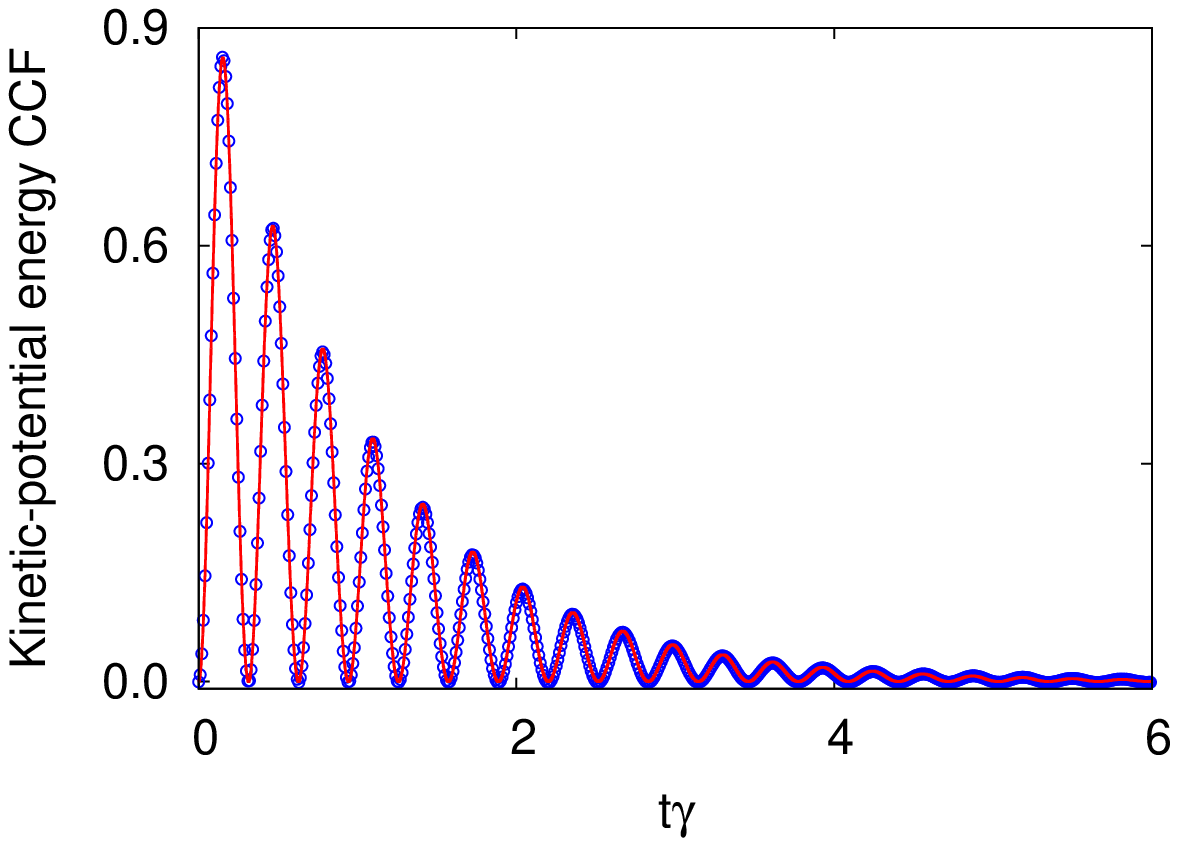}}
\par\end{centering}
\textcolor{black}{\caption{The kinetic-potential energy CCF in the frequency (a) and time (b)
domains for underdamped vibrations with the damping constant $\gamma=0.1\omega_{0}$.
The points and lines represent MD results and analytical solutions,
respectively.\textcolor{black}{{} The functions are normalized according
to Eqs.(\ref{eq:97}) and (\ref{eq:98}).}\label{fig:KU-CCF}}
}
\end{figure}

\textcolor{black}{Selected results are shown in Figs.~\ref{fig:K-ACF}-\ref{fig:KU-CCF}
(for the complete set of figures the reader is referred to the Supplementary
Material \citep{Supplementary-Langevin}), plotting the normalized
correlation functions (\ref{eq:97}) or (\ref{eq:98}) against $\omega/\omega_{0}$
or $t\gamma$ for $\gamma/\omega_{0}=0.1$. Although the spectra only
have physical meaning when $\omega\geq0$, the functions are mathematically
defined in the entire frequency range $(-\infty,\infty)$ and are
shown as such in the figures. The main conclusion of this comparison
is that the analytical solutions accurately match the MD results,
validating the pair-correlation approximation discussed in Appendix
B.}

\section{\textcolor{black}{Application to the problem of temperature fluctuations }}

\textcolor{black}{While fluctuations of extensive parameters, such
as energy, are well-understood, there are controversies regarding
the nature, or even existence \citep{Kittel_Thermal_Physics,Kittel:1973aa,Kittel:1988aa},
of temperature fluctuations in canonical systems \citep{van-Hemmen:2013aa}.
The main source of the controversy is the disparity in the definitions
of certain fundamental concepts, such as entropy and temperature,
in thermodynamics and statistical mechanics. In thermodynamics, temperature
is uniquely defined by the fundamental equation of the substance in
question as the derivative of energy $E$ with respect to entropy
$S$ \citep{Willard_Gibbs,Callen_book_1985,Tisza:1961aa}. For a simple
substance, the fundamental equation has the form $E=E(S,V,N)$, where
$V$ is the system volume and $N$ is the number of particles. By
contrast, the statistical-mechanical definition depends on the adopted
logical structure of the discipline. For example, if temperature of
a canonical system is }\textcolor{black}{\emph{defined}}\textcolor{black}{{}
as the temperature of the thermostat $T_{0}$ (the inverse of $\beta$
in the standard canonical distribution), then of course the very notion
of temperature fluctuations is meaningless \citep{Kittel:1973aa,Kittel:1988aa,Kittel_Thermal_Physics}.
From this point of view, the temperature fluctuation relation
\begin{equation}
\overline{(\Delta T)^{2}}=\dfrac{kT_{0}^{2}}{Nc_{v}^{0}}\label{eq:91}
\end{equation}
derived in the thermodynamic theory of fluctuations \citep{Landau-Lifshitz-Stat-phys,Callen_book_1985,Mishin:2015ab,Hickman_2016}
is the result of a mere manipulation of symbols \citep{Falcioni:2011aa,Kittel:1988aa}.
In Eq.(\ref{eq:91}), $\Delta T=T-T_{0}$, $c_{v}^{0}$ is the constant-volume
specific heat (per particle) at the temperature $T_{0}$, and $k$
is Boltzmann's constant. The system volume and number of particles
are assumed to be fixed. At best, Eq.(\ref{eq:91}) is interpreted
as a rewriting of the known energy fluctuation relation
\begin{equation}
\overline{(\Delta E)^{2}}=NkT_{0}^{2}c_{v}^{0}\label{eq:92}
\end{equation}
by }\textcolor{black}{\emph{formally}}\textcolor{black}{{} defining
the non-equilibrium temperature $T$ as $T\equiv T_{0}+\Delta E/(Nc_{v}^{0})$
\citep{van-Hemmen:2013aa}. This makes $T$ a formal parameter essentially
identical to energy up to units. Other authors suggest that it is
the temperature itself that is not perfectly defined, whereas its
fluctuation is perfectly well defined within the framework of the
statistical estimation theory \citep{Mandelbrot:1989aa,Falcioni:2011aa}.}

\textcolor{black}{By contrast, the thermodynamic theory of fluctuations
\citep{Landau-Lifshitz-Stat-phys,Callen_book_1985,Mishin:2015ab}
endows the non-equilibrium temperature with a physical meaning and
considers its fluctuations as a real physical phenomenon that can
be studied experimentally \citep{Chiu:1992aa}. The theory recognizes
the existence of two different timescales inherent in canonical fluctuations:
the timescale of internal relaxation $t_{r}$ inside the system and
the timescale $\tau_{r}$ of relaxation in the compound system consisting
of the canonical system and the thermostat.}\footnote{\textcolor{black}{An illuminating thermodynamic analysis of system-thermostat
interactions and the role of dissipation by friction in such interactions
can be found in the recent papers \citep{Bizarro:2008aa,Bizarro:aa}.}}\textcolor{black}{{} The two relaxation processes are governed by different
physical mechanisms and, in most cases, $t_{r}\ll\tau_{r}$. Thus,
there is an intermediate timescale $t_{q}$, such that $t_{r}\ll t_{q}\ll\tau_{r}$,
on which the system remains infinitely close to internal equilibrium
without being necessarily in equilibrium with the thermostat. Such
virtually equilibrium states of the canonical system are called }\textcolor{black}{\emph{quasi-equilibrium}}\textcolor{black}{.
On the quasi-equilibrium timescale $t_{q}$, the system can be described
by a fundamental equation, from which its temperature can be found
by
\begin{equation}
T=(\partial E/\partial S)_{V,N}.\label{eq:93}
\end{equation}
}

\textcolor{black}{During the equilibration of a system with a thermostat,
the system goes through a continuum of quasi-equilibrium states. Accordingly,
we can talk about the time evolution of its quasi-equilibrium temperature
$T$ towards $T_{0}$ as the system approaches equilibrium with the
thermostat. Based on the fluctuation-dissipation relation \citep{Landau-Lifshitz-Stat-phys,Nyquist_1928,Onsager1931a,Onsager1931b,Callen_Welton_1951,Kubo:1966aa,Marconia:2008aa},
one can expect that similar quasi-equilibrium states arise during
fluctuations after the system has reached equilibrium with the thermostat.
Such quasi-equilibrium states also have a well-defined temperature
that fluctuates around $T_{0}$. As long as this temperature is properly
defined on the quasi-equilibrium timescale, its fluctuations will
follow Eq.(\ref{eq:91}).}

\textcolor{black}{Similar theories of temperature fluctuations have
been formulated in statistical-mechanical terms by allowing $\beta$
of the canonical distribution to fluctuate away from $\beta_{0}$
of the thermostat \citep{Touchette:aa,Dixit:2015aa}. Such theories
assume, explicitly or implicitly, the existence of timescale separation
and internal equilibration of the system on a certain timescale (which
we call here quasi-equilibrium) with different values of $\beta$.
Such approaches are thus perfectly compatible with ours. }

\textcolor{black}{While Eq.(\ref{eq:93}) provides a thermodynamic
definition of the quasi-equilibrium temperature $T$, in practice
this temperature can be evaluated by utilizing the equipartition relation
and the kinetic energy averaged over the quasi-equilibrium timescale
$t_{q}$. This can be readily done in computer simulations and, in
principle, in experiments measuring a property sensitive to kinetic
energy of the particles. Instead of kinetic energy, other parameters
could be used for computing the temperature \citep{Rugh:1997aa}.
This does not imply an ambiguity in the temperature definition but
rather the possibility of using different ``thermometric properties''
for its evaluation. For example, the potential energy could also be
used for defining the temperature through the appropriate equipartition
relation. A thorough discussion of different definitions of temperature
in statistical mechanics can be found, for example, in \citep{Rugh:1997aa,Prosper:1993aa,Cerino:2015aa}.
This approach obviously assumes ergodicity of the system and classical
dynamics}

\textcolor{black}{The Langevin oscillator offers a simple model that
can illustrate these ideas. Consider the Einstein model of a solid
with a single vibrational frequency $\omega_{0}$. The $3N$ oscillators
describing the atomic vibrations are considered totally decoupled
from each other and only interact with a thermostat. Suppose the latter
is a Langevin thermostat characterized by a damping constant $\gamma$
and a random force $R$ satisfying the fluctuation-dissipation relation
(\ref{eq:17}) for a given thermostat temperature $T_{0}$. The Langevin
thermostat \citep{FrenkelS02} mimics a real thermostat by treating
the atoms as if they were embedded in an artificial viscous medium
composed of much smaller particles. This medium exerts a drag force
as well as a stochastic noise force $R$ that constantly perturbs
the atoms. In this model, each vibrational mode can be represented
by a single Langevin oscillator. The damping time $\tau_{r}=1/\gamma$
sets the timescale of energy exchanges with the thermostat. By contrast
to a real solid wherein internal equilibration requires redistribution
of energy between different vibrational modes by phonon scattering,
in the present model the energy is pumped into or removed from each
oscillator individually. Thus, the internal equilibration timescale
$t_{r}$ is on the order of $1/\omega_{0}$.}\footnote{\textcolor{black}{Perhaps a more accurate measure is the half-period,
$\pi/\omega_{0}$, which is sufficient for the kinetic energy to transform
to potential. But since we are only interested in orders of magnitude,
$1/\omega_{0}$ is a suitable estimate of the relevant timescale.}}\textcolor{black}{{} We assume that the vibrations are quasi-harmonic
and thus $\omega_{0}\gg\gamma$ (underdamped regime). Then $t_{r}\ll\tau_{r}$
and there is a quasi-equilibrium timescale in between on which the
temperature can be defined. }

\textcolor{black}{We have shown above that kinetic energy fluctuations
of an underdamped Langevin oscillator have two components: a fast
component due to transformations between the kinetic and potential
energies during atomic vibrations (period $\pi/\omega_{0}$), and
a slow component due to energy exchanges with the thermostat (timescale
$1/\gamma$). It is the slow component that should be used to calculate
the quasi-equilibrium temperature of the system. The fast component
can be ``filtered out'' by averaging $K$ over several vibration periods.
Alternatively, the same can be achieved by separating the peaks in
the spectrum of the kinetic energy ACF. As was shown above, $\hat{C}_{\Delta K\Delta K}(\omega)$
has two peaks separated by a frequency gap (Figs.~\ref{fig:2}(a)
and ~\ref{fig:K-ACF}(a)). One peak at $\omega=2\omega_{0}$ represents
the kinetic-potential energy exchanges during the vibrations (fast
component) while the other at $\omega=0$ represents the energy exchanges
with the thermostat (slow component). Thus, the separation of the
two timescales can be accomplished by splitting the spectrum in two
Lorentz peaks described by Eqs.(\ref{eq:41}) and (\ref{eq:38}),
respectively. As was shown in Sec.~\ref{sec:KE}, each peak describes
kinetic energy fluctuations with the same variance
\begin{equation}
\overline{(\Delta K)^{2}}=(kT_{0})^{2}/4.\label{eq:94}
\end{equation}
For a solid composed of $3N$ statistically independent oscillators,
we use the low-frequency peak (at $\omega=0$) to obtain
\begin{equation}
\overline{(\Delta K_{\mathrm{solid}})^{2}}=\dfrac{3N(kT_{0})^{2}}{4}.\label{eq:94-1}
\end{equation}
}

\textcolor{black}{We can now identify the quasi-equilibrium temperature
with the equipartition value $T=2K_{\mathrm{solid}}/3Nk$ using the
kinetic energy defined by the low-frequency peak. Inserting this temperature
in Eq.(\ref{eq:94-1}) we have
\begin{equation}
\overline{(\Delta T)^{2}}=\dfrac{T_{0}^{2}}{3N}.\label{eq:95}
\end{equation}
This fluctuation relation matches Eq.(\ref{eq:91}) if $c_{v}^{0}=3k$,
which is exactly the classical specific heat of the solid. We emphasize
that this result was obtained by defining the quasi-equilibrium temperature
using the kinetic energy and without any reference to the behavior
of the total energy during the fluctuations. This is fundamentally
different from the approach mentioned above wherein $T$ is defined
as a formal quantity strictly proportional to $E$. That approach
also leads to Eq.(\ref{eq:95}), except that the latter simply reflects
the temperature definition. As mentioned above, potential energy could
also be used to define the temperature, which would lead to exactly
the same temperature fluctuation (\ref{eq:95}). }

\textcolor{black}{To show that the foregoing derivation of Eq.(\ref{eq:95})
is non-trivial, suppose we ignore the different timescales and define
the temperature from the same equipartition rule but now using }\textcolor{black}{\emph{instantaneous}}\textcolor{black}{{}
values of the kinetic energy, as is often done in MD simulations.
The mean-square fluctuation of this ``instantaneous temperature''
$\tilde{T}$ is obtained by averaging over both timescales or, which
is equivalent, by including both peaks of $\hat{C}_{\Delta K\Delta K}(\omega)$.
As was discussed in Sec.~\ref{sec:KE}, the respective kinetic energy
fluctuation of an oscillator is then $\overline{(\Delta K)^{2}}=(kT_{0})^{2}/2$.
This leads to the temperature fluctuation
\begin{equation}
\overline{(\Delta\tilde{T})^{2}}=\dfrac{2T_{0}^{2}}{3N}.\label{eq:96}
\end{equation}
The specific heat extracted from this fluctuation relation is $c_{v}^{0}=3k/2$,
which is factor of two off. It is only the temperature defined on
the quasi-equilibrium timescale that satisfies the fluctuation relation
(\ref{eq:91}) with the correct specific heat.}

\section{\textcolor{black}{Concluding remarks}}

\textcolor{black}{The main result of this work is the derivation of
the analytical solutions for the energy correlation functions of a
Langevin oscillator. The derivation was enabled by approximating the
quadruple correlation functions by a sum of products of pair correlation
functions as explained in Appendix B. In other words, the derivations
neglect all correlations between stochastic properties beyond pairwise.
The accuracy of this approximation has been validated by comparison
with MD simulations, which were found to be in excellent agreement
with the analytical solutions.}

\textcolor{black}{Given the role of the Langevin oscillator model
in various areas of physics, the results obtained here might be useful
for addressing diverse physics problems involving energy fluctuations
in systems coupled to a thermostat. As one example of possible applications,
we have presented a simple model illustrating the existence and the
meaning of the temperature fluctuations in canonical systems. Temperature
fluctuations is a controversial subject with many conflicting views
published over the past century (see references in \citep{van-Hemmen:2013aa}). }

\textcolor{black}{One of the oldest and, in our opinion, most fruitful
approaches recognizes the existence of quasi-equilibrium states that
arise during canonical fluctuations and exist on a particular timescale
\citep{Landau-Lifshitz-Stat-phys,Callen_book_1985,Mishin:2015ab}.
The temperature calculated on this quasi-equilibrium timescale by
treating the system as if it were equilibrium, is a well-defined physical
property whose fluctuations follow the relation (\ref{eq:91}). By
considering an Einstein solid composed of Langevin oscillators, we
have demonstrated the existence of the quasi-equilibrium timescale
and verified that the temperature computed on this timescale indeed
satisfies Eq.(\ref{eq:91}). Although rather simplistic, this model
captures the essential physics. A more realistic MD study of temperature
fluctuations in a crystalline solid modeled with an accurate many-body
atomistic potential will be published elsewhere \citep{Hickman_2016}.\bigskip{}
}

\textcolor{black}{\emph{Acknowledgments -}}\textcolor{black}{{} This
work was supported by the U.S. Department of Energy, Office of Basic
Energy Sciences, Division of Materials Sciences and Engineering, the
Physical Behavior of Materials Program, through Grant No.~DE-FG02-01ER45871.}

%\textcolor{black}{\bibliographystyle{/Users/ymishin/YURI/Bibliography/ActaMatnew}
%\bibliography{/Users/ymishin/YURI/Bibliography/literat}
%}

\section{\textcolor{black}{Appendix A}}

\textcolor{black}{The Fourier resolution of a function of time $f(t)$
is 
\[
f(t)=\intop_{-\infty}^{\infty}\hat{f}(\omega)e^{i\omega t}d\omega,
\]
with the Fourier amplitude
\[
\hat{f}(\omega)=\dfrac{1}{2\pi}\intop_{-\infty}^{\infty}f(t)e^{-i\omega t}dt.
\]
The Fourier transform of a product of two functions is the convolution
of their Fourier transforms and vise versa: if $R(t)=f(t)g(t)$, then
\begin{equation}
\hat{R}(\omega)=\intop_{-\infty}^{\infty}\hat{f}(\omega-\omega^{\prime})\hat{g}(\omega^{\prime})d\omega^{\prime},\label{eq:A6}
\end{equation}
and if $\hat{R}(\omega)=\hat{f}(\omega)\hat{g}(\omega)$, then
\begin{equation}
R(t)=\dfrac{1}{2\pi}\intop_{-\infty}^{\infty}f(t-t^{\prime})g(t^{\prime})dt^{\prime}.\label{eq:A7}
\end{equation}
Calculations involving Dirac's delta-function utilize the relations
$\hat{1}(\omega)=\delta(\omega)$ and $\hat{\delta}(\omega)=1/2\pi$.
Spectral calculations often employ the residue theorem of complex
analysis. The residues can be found analytically or with the help
of the }\textcolor{black}{\emph{Wolfram Mathematica$^{\circledR}$}}\textcolor{black}{{}
function }\textsf{\textcolor{black}{Residue{[}{]}}}\textcolor{black}{. }

\textcolor{black}{The pair correlation function of two (generally,
complex) stochastic variables $y$ and $z$ is defined by 
\begin{equation}
C_{yz}(t)=\overline{y(t^{\prime})z(t^{\prime}+t)}=\overline{y(0)z(t)},\label{eq:A17}
\end{equation}
where we assumed that the process is stationary and thus independent
of the initial time $t^{\prime}$. Obviously, $C_{yz}(t)=C_{zy}(-t)$.
The Fourier transform 
\begin{equation}
\hat{C}_{yz}(\omega)=\dfrac{1}{2\pi}\intop_{-\infty}^{\infty}C_{yz}(t)e^{-i\omega t}dt\label{eq:A20}
\end{equation}
has the property $\hat{C}_{yz}(\omega)=\hat{C}_{zy}(-\omega)$. The
inverse transformation recovers $C_{yz}(t)$:
\begin{equation}
C_{yz}(t)=\intop_{-\infty}^{\infty}\hat{C}_{yz}(\omega)e^{i\omega t}d\omega.\label{eq:A21}
\end{equation}
Taking $t=0$ we obtain
\begin{equation}
\overline{yz}=C_{yz}(0)=\intop_{-\infty}^{\infty}\hat{C}_{yz}(\omega)d\omega.\label{eq:A23}
\end{equation}
It can be shown that 
\begin{equation}
\overline{\hat{y}(\omega)\hat{z}(\omega^{\prime})}=\delta(\omega+\omega^{\prime})\hat{C}_{yz}(\omega^{\prime})=\delta(\omega+\omega^{\prime})\hat{C}_{yz}(-\omega).\label{eq:A19}
\end{equation}
Integrating the last equation with respect to $\omega^{\prime}$ we
find
\begin{equation}
\hat{C}_{yz}(\omega)=\intop_{-\infty}^{\infty}\overline{\hat{y}(\omega^{\prime})\hat{z}(\omega)}d\omega^{\prime}.\label{eq:A24}
\end{equation}
}

\textcolor{black}{In the particular case when $y(t)\equiv z(t)$,
we obtain the autocorrelation function (ACF)
\begin{equation}
C_{yy}(t)=\overline{y(0)y(t)}\label{eq:A25}
\end{equation}
and its Fourier transform $\hat{C}_{yy}(\omega)$. Both functions
are symmetric: $C_{yy}(t)=C_{yy}(-t)$ and $\hat{C}_{yy}(\omega)=\hat{C}_{yy}(-\omega)$.
Equations (\ref{eq:A23}), (\ref{eq:A19}) and (\ref{eq:A24}) become,
respectively,
\begin{equation}
\overline{y^{2}}=\intop_{-\infty}^{\infty}\hat{C}_{yy}(\omega)d\omega,\label{eq:A27}
\end{equation}
\begin{equation}
\overline{\hat{y}(\omega)\hat{y}(\omega^{\prime})}=\delta(\omega+\omega^{\prime})\hat{C}_{yy}(\omega),\label{eq:A28}
\end{equation}
\begin{equation}
\hat{C}_{yy}(\omega)=\intop_{-\infty}^{\infty}\overline{\hat{y}(\omega^{\prime})\hat{y}(\omega)}d\omega^{\prime}.\label{eq:A29}
\end{equation}
}

\section{\textcolor{black}{Appendix B}}

\textcolor{black}{For two (generally, complex) stochastic properties
$a(t)$ and $b(t)$, let us evaluate the ACF of $F(t)=a(t)b(t)$ relative
to its average value $\overline{F}=\overline{ab}$. Denoting $\Delta F=F-\overline{F}$,
we have
\begin{equation}
C_{\Delta F\Delta F}(t)=C_{FF}(t)-(\overline{ab})^{2}.\label{eq:B1}
\end{equation}
It will suffice to find the Fourier transform $\hat{C}_{\Delta F\Delta F}(\omega)$,
which can be then inverted to $C_{\Delta F\Delta F}(t)$.}

\textcolor{black}{By the product rule of the Fourier transformation,
\begin{equation}
\hat{F}(\omega)=\intop_{-\infty}^{\infty}\hat{a}(\omega^{\prime})\hat{b}(\omega-\omega^{\prime})d\omega^{\prime}.\label{eq:B2}
\end{equation}
Applying this rule twice and averaging over the ensemble we obtain
\begin{equation}
\overline{\hat{F}(\omega)\hat{F}(\omega^{\prime})}=\intop_{-\infty}^{\infty}\intop_{-\infty}^{\infty}\overline{\hat{a}(\omega^{\prime\prime})\hat{b}(\omega-\omega^{\prime\prime})\hat{a}(\omega^{\prime\prime\prime})\hat{b}(\omega^{\prime}-\omega^{\prime\prime\prime})}d\omega^{\prime\prime}d\omega^{\prime\prime\prime}.\label{eq:B3}
\end{equation}
We will assume that the quadruple correlation function appearing in
this equation can be broken into a sum of products of pair correlation
functions. Only three distinct products can be formed, which are obtained
by permutations of the $\hat{a}$'s and $\hat{b}$'s:
\[
\overline{\hat{a}(\omega^{\prime\prime})\hat{b}(\omega-\omega^{\prime\prime})}\:\:\overline{\hat{a}(\omega^{\prime\prime\prime})\hat{b}(\omega^{\prime}-\omega^{\prime\prime\prime})},
\]
\[
\overline{\hat{a}(\omega^{\prime\prime})\hat{a}(\omega^{\prime\prime\prime})}\:\:\overline{\hat{b}(\omega-\omega^{\prime\prime})\hat{b}(\omega^{\prime}-\omega^{\prime\prime\prime})},
\]
\[
\overline{\hat{a}(\omega^{\prime\prime})\hat{b}(\omega^{\prime}-\omega^{\prime\prime\prime})}\:\:\overline{\hat{a}(\omega^{\prime\prime\prime})\hat{b}(\omega-\omega^{\prime\prime})}.
\]
Applying Eq.(\ref{eq:A19}), these functions become, respectively,
\[
\delta(\omega)\delta(\omega^{\prime})\hat{C}_{ab}(\omega-\omega^{\prime\prime})\hat{C}_{ab}(\omega^{\prime}-\omega^{\prime\prime\prime}),
\]
\[
\delta(\omega^{\prime\prime}+\omega^{\prime\prime\prime})\delta(\omega+\omega^{\prime}-\omega^{\prime\prime}-\omega^{\prime\prime\prime})\hat{C}_{aa}(\omega^{\prime\prime\prime})\hat{C}_{bb}(\omega^{\prime}-\omega^{\prime\prime\prime}),
\]
\[
\delta(\omega^{\prime}+\omega^{\prime\prime}-\omega^{\prime\prime\prime})\delta(\omega-\omega^{\prime\prime}+\omega^{\prime\prime\prime})\hat{C}_{ab}(\omega^{\prime}-\omega^{\prime\prime\prime})\hat{C}_{ab}(\omega-\omega^{\prime\prime}).
\]
Inserting their sum in Eq.(\ref{eq:B3}) we obtain 
\begin{eqnarray}
\overline{\hat{F}(\omega)\hat{F}(\omega^{\prime})} & = & \delta(\omega)\delta(\omega^{\prime})\left(\intop_{-\infty}^{\infty}\hat{C}_{ab}(-\omega^{\prime\prime})d\omega^{\prime\prime}\right)\left(\intop_{-\infty}^{\infty}\hat{C}_{ab}(-\omega^{\prime\prime\prime})d\omega^{\prime\prime\prime}\right)\nonumber \\
 & + & \delta(\omega+\omega^{\prime})\intop_{-\infty}^{\infty}\hat{C}_{aa}(\omega^{\prime\prime\prime})\hat{C}_{bb}(\omega^{\prime}-\omega^{\prime\prime\prime})d\omega^{\prime\prime\prime}\nonumber \\
 & + & \delta(\omega+\omega^{\prime})\intop_{-\infty}^{\infty}\hat{C}_{ba}(\omega^{\prime\prime})\hat{C}_{ab}(\omega-\omega^{\prime\prime})d\omega^{\prime\prime}.\label{eq:B4}
\end{eqnarray}
By Eq.(\ref{eq:A23}), the first line gives $\delta(\omega)\delta(\omega^{\prime})(\overline{ab})^{2}$. }

\textcolor{black}{Integrating Eq.(\ref{eq:B4}) with respect to $\omega^{\prime}$
and applying Eq.(\ref{eq:A29}) we obtain
\begin{equation}
\hat{C}_{FF}(\omega)=\delta(\omega)(\overline{ab})^{2}+\intop_{-\infty}^{\infty}\hat{C}_{aa}(\omega^{\prime\prime})\hat{C}_{bb}(\omega-\omega^{\prime\prime})d\omega^{\prime\prime}+\intop_{-\infty}^{\infty}\hat{C}_{ba}(\omega^{\prime\prime})\hat{C}_{ab}(\omega-\omega^{\prime\prime})d\omega^{\prime\prime}.\label{eq:B5}
\end{equation}
On the other hand, the Fourier transform of Eq.(\ref{eq:B1}) is 
\begin{equation}
\hat{C}_{\Delta F\Delta F}(\omega)=\hat{C}_{FF}(\omega)-\delta(\omega)(\overline{ab})^{2}.\label{eq:B6}
\end{equation}
Comparing Eqs.(\ref{eq:B5}) and (\ref{eq:B6}), we obtain
\[
\hat{C}_{\Delta F\Delta F}(\omega)=\intop_{-\infty}^{\infty}\hat{C}_{aa}(\omega^{\prime})\hat{C}_{bb}(\omega-\omega^{\prime})d\omega^{\prime}+\intop_{-\infty}^{\infty}\hat{C}_{ba}(\omega^{\prime})\hat{C}_{ab}(\omega-\omega^{\prime})d\omega^{\prime}.
\]
}

\textcolor{black}{Next, we will take the same stochastic properties
$a(t)$ and $b(t)$, form two new properties $G(t)=a^{2}(t)$ and
$H(t)=b^{2}(t)$, and evaluate the cross-correlation function (CCF)
$C_{\Delta G\Delta H}(t)$, where $\Delta G=G-\overline{G}$ and $\Delta H=H-\overline{H}$.
It will suffice to find the Fourier transform 
\begin{equation}
\hat{C}_{\Delta G\Delta H}(\omega)=\hat{C}_{GH}(\omega)-\delta(\omega)\overline{G}\:\overline{H}.\label{eq:B1-1}
\end{equation}
Applying the product rule of Fourier transformations we have
\begin{equation}
\overline{\hat{G}(\omega)\hat{H}(\omega^{\prime})}=\intop_{-\infty}^{\infty}\intop_{-\infty}^{\infty}\overline{\hat{a}(\omega^{\prime\prime})\hat{a}(\omega-\omega^{\prime\prime})\hat{b}(\omega^{\prime\prime\prime})\hat{b}(\omega^{\prime}-\omega^{\prime\prime\prime})}d\omega^{\prime\prime}d\omega^{\prime\prime\prime}.\label{eq:B3-1}
\end{equation}
As above, we break the quadruple correlation function into a sum of
products of pair correlation functions. The three distinct products
are
\[
\overline{\hat{a}(\omega^{\prime\prime})\hat{a}(\omega-\omega^{\prime\prime})}\:\:\overline{\hat{b}(\omega^{\prime\prime\prime})\hat{b}(\omega^{\prime}-\omega^{\prime\prime\prime})}=\delta(\omega)\delta(\omega^{\prime})\hat{C}_{aa}(\omega^{\prime\prime})\hat{C}_{bb}(\omega^{\prime\prime\prime}),
\]
\[
\overline{\hat{a}(\omega^{\prime\prime})\hat{b}(\omega^{\prime\prime\prime})}\:\:\overline{\hat{a}(\omega-\omega^{\prime\prime})\hat{b}(\omega^{\prime}-\omega^{\prime\prime\prime})}=\delta(\omega^{\prime\prime}+\omega^{\prime\prime\prime})\delta(\omega+\omega^{\prime}-\omega^{\prime\prime}-\omega^{\prime\prime\prime})\hat{C}_{ab}(\omega^{\prime\prime\prime})\hat{C}_{ab}(\omega^{\prime}-\omega^{\prime\prime\prime}),
\]
\[
\overline{\hat{a}(\omega^{\prime\prime})\hat{b}(\omega^{\prime}-\omega^{\prime\prime\prime})}\:\:\overline{\hat{b}(\omega^{\prime\prime\prime})\hat{a}(\omega-\omega^{\prime\prime})}=\delta(\omega^{\prime}+\omega^{\prime\prime}-\omega^{\prime\prime\prime})\delta(\omega-\omega^{\prime\prime}+\omega^{\prime\prime\prime})\hat{C}_{ab}(\omega^{\prime}-\omega^{\prime\prime\prime})\hat{C}_{ab}(\omega^{\prime\prime\prime}),
\]
Inserting the sum of these terms in Eq.(\ref{eq:B3-1}) we have 
\begin{eqnarray*}
\overline{\hat{G}(\omega)\hat{H}(\omega^{\prime})} & = & \delta(\omega)\delta(\omega^{\prime})\left(\intop_{-\infty}^{\infty}\hat{C}_{aa}(\omega^{\prime\prime})d\omega^{\prime\prime}\right)\left(\intop_{-\infty}^{\infty}\hat{C}_{bb}(\omega^{\prime\prime\prime})d\omega^{\prime\prime\prime}\right)\\
 & + & 2\delta(\omega+\omega^{\prime})\intop_{-\infty}^{\infty}\hat{C}_{ab}(\omega^{\prime\prime\prime})\hat{C}_{ab}(\omega^{\prime}-\omega^{\prime\prime\prime})d\omega^{\prime\prime\prime}\\
 & = & \delta(\omega)\delta(\omega^{\prime})\overline{G}\:\overline{H}+2\delta(\omega+\omega^{\prime})\intop_{-\infty}^{\infty}\hat{C}_{ab}(\omega^{\prime\prime\prime})\hat{C}_{ab}(\omega^{\prime}-\omega^{\prime\prime\prime})d\omega^{\prime\prime\prime}.
\end{eqnarray*}
Comparing this equation with Eq.(\ref{eq:A19}) and applying Eq.(\ref{eq:B1-1}),
we obtain
\begin{equation}
\hat{C}_{\Delta G\Delta H}(\omega)=2\intop_{-\infty}^{\infty}\hat{C}_{ab}(\omega^{\prime})\hat{C}_{ab}(\omega-\omega^{\prime})d\omega^{\prime}.\label{eq:B7-1}
\end{equation}
}

\textcolor{black}{The foregoing results can be summarized as the following
statement:}

\textcolor{black}{\emph{If only pair correlations are taken into account,
then for any two stochastic properties $a(t)$ and $b(t)$, }}\textcolor{black}{
\begin{equation}
\hat{C}_{\Delta F\Delta F}(\omega)=\intop_{-\infty}^{\infty}\hat{C}_{aa}(\omega^{\prime})\hat{C}_{bb}(\omega-\omega^{\prime})d\omega^{\prime}+\intop_{-\infty}^{\infty}\hat{C}_{ba}(\omega^{\prime})\hat{C}_{ab}(\omega-\omega^{\prime})d\omega^{\prime},\label{eq:B7}
\end{equation}
\begin{equation}
\hat{C}_{\Delta G\Delta H}(\omega)=2\intop_{-\infty}^{\infty}\hat{C}_{ab}(\omega^{\prime})\hat{C}_{ab}(\omega-\omega^{\prime})d\omega^{\prime},\label{eq:B7-1-1}
\end{equation}
}\textcolor{black}{\emph{where $F(t)=a(t)b(t)$, }}\textcolor{black}{$G(t)=a^{2}(t)$
}\textcolor{black}{\emph{and}}\textcolor{black}{{} $H(t)=b^{2}(t)$.}

\textcolor{black}{In the particular case when $a(t)\equiv b(t)$,
we have $F(t)=a^{2}(t)$ and Eq.(\ref{eq:B7}) gives
\begin{equation}
\hat{C}_{\Delta F\Delta F}(\omega)=2\intop_{-\infty}^{\infty}\hat{C}_{aa}(\omega^{\prime})\hat{C}_{aa}(\omega-\omega^{\prime})d\omega^{\prime}.\label{eq:B9}
\end{equation}
}

\textcolor{black}{\newpage{}\clearpage{}}

\textcolor{black}{}

\textcolor{black}{}

\textcolor{black}{}

\textcolor{black}{}

\textcolor{black}{}

\textcolor{black}{}

\end{document}